\documentclass[preprint]{aastex}
\usepackage{csvsimple}
\usepackage{adjustbox}
\usepackage{amsmath}

\shorttitle{Near-Infrared Spectra of KH 15D}
\shortauthors{Arulanantham et al.}

\begin{document}

\title{Untangling the Near-IR Spectral Features in the Protoplanetary Environment of KH 15D}

\author{\thanks{Currently at University of Colorado, Boulder (nicole.arulanantham@colorado.edu)} Nicole A. Arulanantham, William Herbst, Martha S. Gilmore, and P. Wilson Cauley}
\affil{Astronomy Department, Wesleyan University, Middletown, CT 06459}
\author{S.K. Leggett}
\affil{Gemini Observatory (North), Hilo, HI 96720}

\begin{abstract}

We report on Gemini/GNIRS observations of the binary T Tauri system V582 Mon (KH 15D) at three orbital phases. These spectra allow us to untangle five components of the system: the photosphere and magnetosphere of star B, the jet, scattering properties of the ring material, and excess near-IR radiation previously attributed to a possible self-luminous planet. We confirm an early-K subgiant classification for star B and show that the magnetospheric He I emission line is variable, possibly indicating increased mass accretion at certain times. As expected, the H$_2$ emission features associated with the inner part of the jet show no variation with orbital phase. We show that the reflectance spectrum for the scattered light has a distinctive blue slope and spectral features consistent with scattering and absorption by a mixture of water and methane ice grains in the 1-50 $\mu$m size range. This suggests that the methane frost line is closer than $\sim$5 AU in this system, requiring that the grains be shielded from direct radiation. After correcting for features from the scattered light, jet, magnetosphere, and photosphere, we confirm the presence of leftover near-IR light from an additional source, detectable near minimum brightness. A spectral emission feature matching the model spectrum of a 10 M$_{J}$, 1 Myr old planet is found in the excess flux, but other expected features from this model are not seen. Our observations, therefore, tentatively support the picture that a luminous planet is present within the system, although they cannot yet be considered definitive. 

\end{abstract}

\keywords{stars: individual (KH15D) --- stars: pre-main sequence --- protoplanetary disks --- clusters: individual (NGC 2264)}

\section{Introduction}

KH 15D is a $\sim$1-3 Myr T Tauri binary system located at a distance of 760 pc in the open cluster NGC 2264. Its stellar components are spectrally classified as K6/K7 (star A; \cite{Hamilton2001}) and K1 (star B; \cite{Capelo2012}). The object was initially thought to be a single variable star, but extensive monitoring revealed unusual properties, in particular a 48 day period of variability with $\sim$4 mag changes in brightness \citep{Kearns1998}. The long-term \emph{I} band light curve shows that the depth of the brightness variations gradually increased between 1995 and 2005, and the peak brightness of the system decreased between 2006 and 2010 \citep{Hamilton2001, Hamilton2005}. These features have been attributed to occultations by a slightly inclined, rigidly precessing circumbinary ring, which causes observed eclipses on the 48.37 day binary orbital period. Before 2010, the leading edge of the ring was precessing across the orbit of star A, while star B was completely occulted at all phases. Both stars were hidden from view between 2010 and 2012, although the system could still be detected in scattered light. KH 15D has steadily been re-brightening since 2012, as the trailing edge of the ring uncovers the orbit of star B \citep{Capelo2012, Windemuth2014, A16}. 

The circumbinary ring must have sharp inner and outer edges in order to precess in the manner that produces the observed features in the long-term light curve \citep{CM2004}. The inner radius is truncated at $\sim$1 AU by tidal forces from the eccentric central binary ($e \sim 0.6$;  \cite{Winn2006}). \cite{CM2004} proposed a planet as a mechanism to shepherd the outer edge of the ring. Although the presence of a planet in the system has not been spectroscopically confirmed, color excesses detected in \emph{I-J} and \emph{I-H} near minimum light are consistent with the expected flux from a 10 $M_J$ planet at an age of $\sim$1 Myr \citep{Windemuth2014}. 

The distance to the system makes it nearly impossible to obtain high resolution radial velocity measurements that could confirm the presence of a planet via Doppler spectroscopy. However, the optically thick circumbinary ring acts as a natural coronagraph during eclipse, potentially allowing spectral features of the planet to be detected against the background scattered starlight. Subtracting the stellar features from spectra taken near mid-eclipse should help isolate the signature of any non-stellar component present in the system. Models from \citet{Spiegel2012} predict that the likely temperature of a 1 Myr, 10 $M_J$ planet is around 2000 K, so observations in the near-IR would be required to detect its features.

\begin{figure}
\centering
\includegraphics[width=1.0\linewidth]
{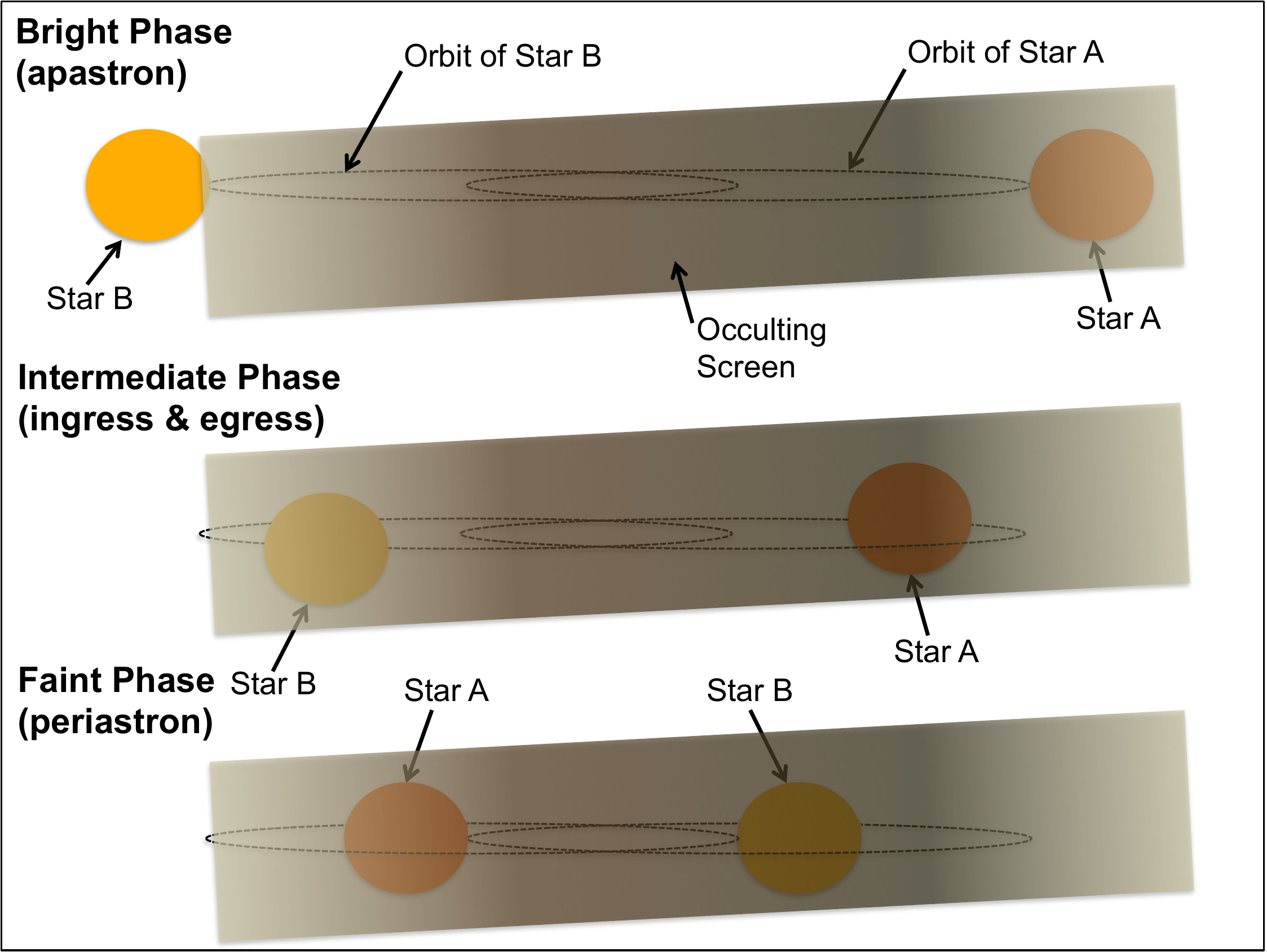}
\caption{We have obtained near-infrared spectra of KH 15D at three orbital phases. The ``bright phase" spectrum was taken near apastron, when star B was directly visible. An ``intermediate phase" spectrum was also acquired when star B was just below the ring edge. Finally, two spectra were collected during different ``faint phases," when the stars were near periastron and the system was at its faintest.}
\label{phase_cartoon}
\end{figure}

When both stars of the binary are fully obscured by the ring, we detect the system only by its scattered radiation. Because of our viewing angle, this is likely to be forward scattered light \citep{Silvia2008}, primarily from the brighter component, star B. The strength of the scattered component also varies with orbital phase and is strongest just after the full occultation of star B, i.e. just after full ``star set". We have, therefore, obtained a set of Gemini/GNIRS spectra at three orbital phases (see Figure \ref{phase_cartoon}), which we label ``bright" (much of the photosphere of star B fully exposed), ``intermediate" (just after full occultation of star B, when the scattered light component is strongest), and ``faint" (when the scattered light component is at its weakest; two spectra, separated by one orbital cycle, were obtained during the faint phase). Here we present and analyze information obtained on the following components of the system: the photosphere of star B, the magnetosphere of star B, the jet, the scattered light, and the putative giant planet.

\section{Observations \& Preliminary Reductions}

Near-infrared spectra of KH 15D, spanning the wavelength range 0.8-2.5 $\mu$m, were obtained on four nights with the Gemini Near-Infrared Spectrograph (GNIRS) at Gemini-North Observatory (see Table \ref{spectraprops}). One spectrum was collected during the bright phase on UT Apr. 3, 2013, when star B was close to apastron, although still partially obscured. A second spectrum was acquired at intermediate phase when star B was fully occulted but close to the edge of the screen (UT Feb. 6, 2013). The last two spectra were taken during the faint phase (UT Nov. 7, 2013; UT Dec. 25, 2013), when both stars were completely hidden and near periastron, as far from the occulting edges of the ring as they get. Direct starlight is fully attenuated during the intermediate and faint phase because of the high opacity of the ring material at these wavelengths, and scattered light is minimized at the faint phase because both stars are well below the projected edges of the occulting disk. 

The spectra were reduced using the standard procedure for GNIRS cross-dispersed data. The XD\_G0526 filter was used for all four observations along with the short blue camera and 32 lines/mm grating. A slit width of 0.30$'$$'$ was used for the bright and intermediate phase observations, corresponding to a resolving power of $R \sim 1800$. However, the slit width had to be increased to 0.45$'$$'$ for the faint phase observations, resulting in lower resolution for those two spectra. An argon arc lamp was used to calibrate the wavelength scale for all four spectra. Telluric atmosphere removal and flux calibration were done using reference stars that were observed at the same time the spectra were collected (see Table \ref{telluric}). 

Observing conditions were partly cloudy on the night that the November faint phase spectrum was obtained. To compensate, KH 15D was observed for 1.8 hours during this observing run (compared to 0.8 hours on source when the December spectrum was obtained), resulting in similar S/N between the two spectra. However, variable seeing on the November night could have impacted observations of the standard stars. The resulting flux calibration is somewhat more uncertain than the flux calibration for the December spectrum, although both are accurate to $\sim$20$\%$. 

\begin{deluxetable}{cccccc}
\tabletypesize{\scriptsize}
\tablewidth{0 pt}
\tablecaption{Properties of Gemini Spectra \label{spectraprops}}
\tablehead{
\colhead{UT Date} & \colhead{UT Start Time} & \colhead{Exposure Time} & \colhead{Slit Width} & \colhead{Position Angle} & \colhead{Resolving Power} \\ \vspace{-0.2cm}
 & \colhead{(hh:mm:ss)} & \colhead{(s)} & \colhead{(arcsec)} & \colhead{(deg)} & \colhead{$\left(\lambda / \Delta \lambda \right)$} \\
}
\startdata

Feb. 6, 2013 & 06:31:38.3 & 300 & 0.30 & 90.0 & 1800  \\
Apr. 3, 2013 & 06:15:52.4 & 90 & 0.30 & 90.0 & 1800 \\
Nov. 7, 2013 & 12:56:43.6 & 300 & 0.45 & 90.0 & $<$ 1800  \\
Dec. 25, 2013 & 12:21:56.9 & 300 & 0.45 & 90.0 & $<$ 1800 \\
\enddata
\end{deluxetable}

\begin{deluxetable}{cccccc}
\tabletypesize{\scriptsize}
\tablewidth{0 pt}
\tablecaption{Telluric Standards from \cite{2MASS} \label{telluric}}
\tablehead{
\colhead{Identifier} & \colhead{Date Observed} & \colhead{Spectral Type} & \colhead{\emph{J}} & \colhead{\emph{H}} & \colhead{\emph{K}} \\ \vspace{-0.2 cm}
 &  \colhead{(dd/mm/yy)} & & \colhead{(mag)} & \colhead{(mag)} & \colhead{(mag)} \\
}
\startdata
HD 37650 & Feb. 6, 2013 &  A & 8.601 $\pm$ 0.027 & 8.579 $\pm$ 0.018 & 8.561 $\pm$ 0.018 \\
HD 60778 & Feb. 6, 2013 & A1V & 8.746 $\pm$ 0.019 & 8.662 $\pm$ 0.031 & 8.666 $\pm$ 0.023 \\ 
HD 52431 & Apr. 3, 2013 & F5V & 6.722 $\pm$ 0.021 & 6.548 $\pm$ 0.047 & 6.496 $\pm$ 0.018 \\
HD 65158 & Nov. 7, 2013 & A0V & 7.093 $\pm$ 0.023 & 7.134 $\pm$ 0.049 & 7.105 $\pm$ 0.017 \\
HD 33140 & Nov. 7, 2013; Dec. 25, 2013 & F2V & 8.435 $\pm$ 0.026 & 8.269 $\pm$ 0.036 & 8.207 $\pm$ 0.023 \\
\enddata
\end{deluxetable}

Nightly ANDICAM optical and near-infrared observations of KH 15D were also in progress at the time the GNIRS spectra were acquired. Table \ref{datematch} lists the \emph{VRIJHK} magnitudes of the system on the nights that the spectra were obtained. The system was observed with ANDICAM within 1 JD of the April 3rd, November 7th, and December 25th Gemini observing runs in 2013. However, the closest data points to the intermediate phase observation on February 6th were collected almost 2 days after the spectrum. In order to obtain a good estimate of the brightness of the star at the time that spectrum was acquired, we took the following steps to interpolate the light curves. The cycle-to-cycle variations in the shape of the light curves were small at this time, so data from the two preceding cycles as well as the two following cycles were phase-folded to fill in gaps. The data were then fit with a univariate spline interpolation over a range of $\pm$0.10 in phase, and errors were estimated from the goodness of fit. The resulting magnitudes are denoted by asterisks in Table \ref{datematch}. 

\begin{deluxetable}{ccccccccc}
\tabletypesize{\scriptsize}
\tablewidth{0 pt}
\tablecaption{Brightness of KH 15D at time of GNIRS Observations \label{datematch}}
\tablehead{
\colhead{UT Date} & \colhead{Julian Date} & \colhead{Phase} & \colhead{$V$} & \colhead{$R$} & \colhead{$I$} & \colhead{$J$} & \colhead{$H$} & \colhead{$K$} \\ \vspace{-0.2 cm}
 & \colhead{(2456000.0)}  &  & \colhead{(mag)} & \colhead{(mag)} & \colhead{(mag)} & \colhead{(mag)} & \colhead{(mag)} & \colhead{(mag)} \\
}
\startdata
Feb. 6, 2013\tablenotemark{*} & 329.78 & 0.24 & 18.8 $\pm$ 0.1 & 17.9 $\pm$ 0.1 & 17.6 $\pm$ 0.1 & 16.6 $\pm$ 0.1 & 15.9 $\pm$ 0.1 & 16.4 $\pm$ 0.1 \\
Apr. 3, 2013 & 385.77 & 0.39 & 16.40 $\pm$ 0.02 & 15.66 $\pm$ 0.02 & 15.25 $\pm$ 0.02 & 14.14 $\pm$ 0.02 & 13.43 $\pm$ 0.02 & 13.29 $\pm$ 0.02 \\
Nov. 7, 2013 & 603.61 & 0.90 & 19.5 $\pm$ 0.1 & 18.8 $\pm$ 0.1 & 18.03 $\pm$ 0.07 & 17.22 $\pm$ 0.08 & 16.8 $\pm$ 0.1 & 17.4 $\pm$ 0.7 \\
Dec. 25, 2013 & 651.55 & 0.89 & 19.5 $\pm$ 0.2 & 18.32 $\pm$ 0.08 & 17.96 $\pm$ 0.06 & 17.6 $\pm$ 0.1 & 16.76 $\pm$ 0.08 & -\tablenotemark{**} \\
\enddata
\tablenotetext{*}{interpolated magnitudes (see text)}
\tablenotetext{**}{magnitude is unknown}
\end{deluxetable}

\section{Results}

\begin{figure}
\includegraphics[width=0.9\linewidth]
{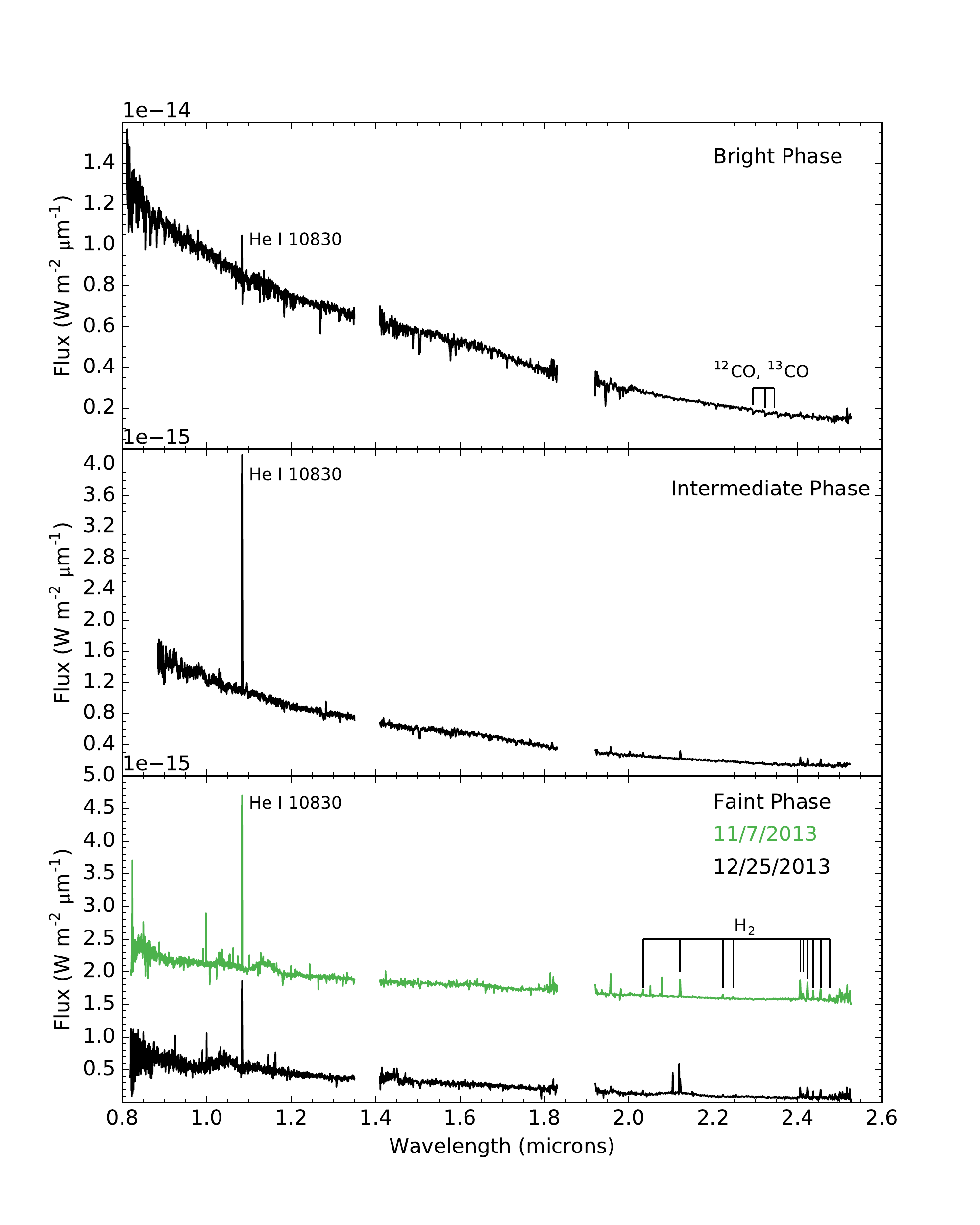}
\caption{The top panel shows the spectrum of KH 15D during its ``bright" phase, when the amount of direct starlight was greatest. The middle spectrum (``intermediate" phase) was taken when star B was just below the edge of the ring. Both spectra in the bottom panel were obtained during ``faint" phases from two different cycles, when both stars were near periastron and the contribution from starlight was minimized. The spectrum from November has been offset by $1.5 \times 10^{-15}$ W m$^{-2}$ $\mu$m$^{-1}$ for comparison to the data from December.}
\label{fullspectrum}
\end{figure}

The four sets of reduced spectra are shown in Figure \ref{fullspectrum}, where the November faint phase spectrum has been shifted to higher flux values for ease of comparison to the December faint spectrum. Some of the most prominent features are the CO absorptions at $\sim$2-2.5 microns in the bright spectrum and the H$_2$ lines in the intermediate and faint spectra, along with strong He I emission at all three phases. Broad features in the continua of both the intermediate and faint phase spectra are also distinguishable. 

According to our current picture of the system, the bright spectrum corresponds to an eclipse phase when some fraction of star B is directly visible \citep{Capelo2012, Windemuth2014}. At this phase, the star itself produces the dominant features in the spectrum. By the intermediate phase, the entire star has been occulted by the ring, although it is still very close to its edge. The ring material is optically thick at the observed wavelengths (0.8-2.5 $\mu$m) \citep{Windemuth2014, A16}, but forward scattering effects allow starlight to escape the system \citep{Agol2004, Silvia2008}. Since the photons detected at this phase have scattered off of dust in the ring, they could potentially carry near-IR reflectance features of the ring itself. The forward scattered light is minimized during the faint phase, when the stars are close to periastron. The faint phase spectra therefore provide the best opportunity to detect a luminous third body in the system.

In order to accurately extract the spectrum of a planet from these data, the features that originate from other components of the system must first be characterized. We have identified seven distinct physical zones that can produce features in the KH 15D spectra: the photospheres of star A and star B, the magnetospheres of both stars, the circumbinary ring (which potentially adds reflectance features, increasing the system's brightness by $\sim$0.5 mag), the inner jet, and a putative young giant planet. \citet{Windemuth2014} predicted that star B would have a magnitude of $I = 14.19 \pm 0.06$ when completely unobscured, making it $\sim$1.3 times as bright as star A $\left(I = 14.47 \pm 0.04 \, \text{mag}\right)$. Star A is completely hidden by the ring at all phases observed here, and we expect that its largest contribution to the total system light comes when it makes its closest approach to the occulting (trailing) edge of the ring (near periastron). At the time that the leading edge of the screen first covered all of star A's orbit, the system's brightness dropped as low as $I \sim 19$ mag during periastron. This is $\sim$1 mag dimmer than the faint phase brightness measured during the observations presented here, so we assume that the photosphere and magnetosphere of star A made negligible contributions to the total system light at this time. An analysis of each of the remaining five regions is presented below.

\subsection{Photosphere of Star B}

\begin{figure}
\centering
\includegraphics[width=0.8\linewidth]
{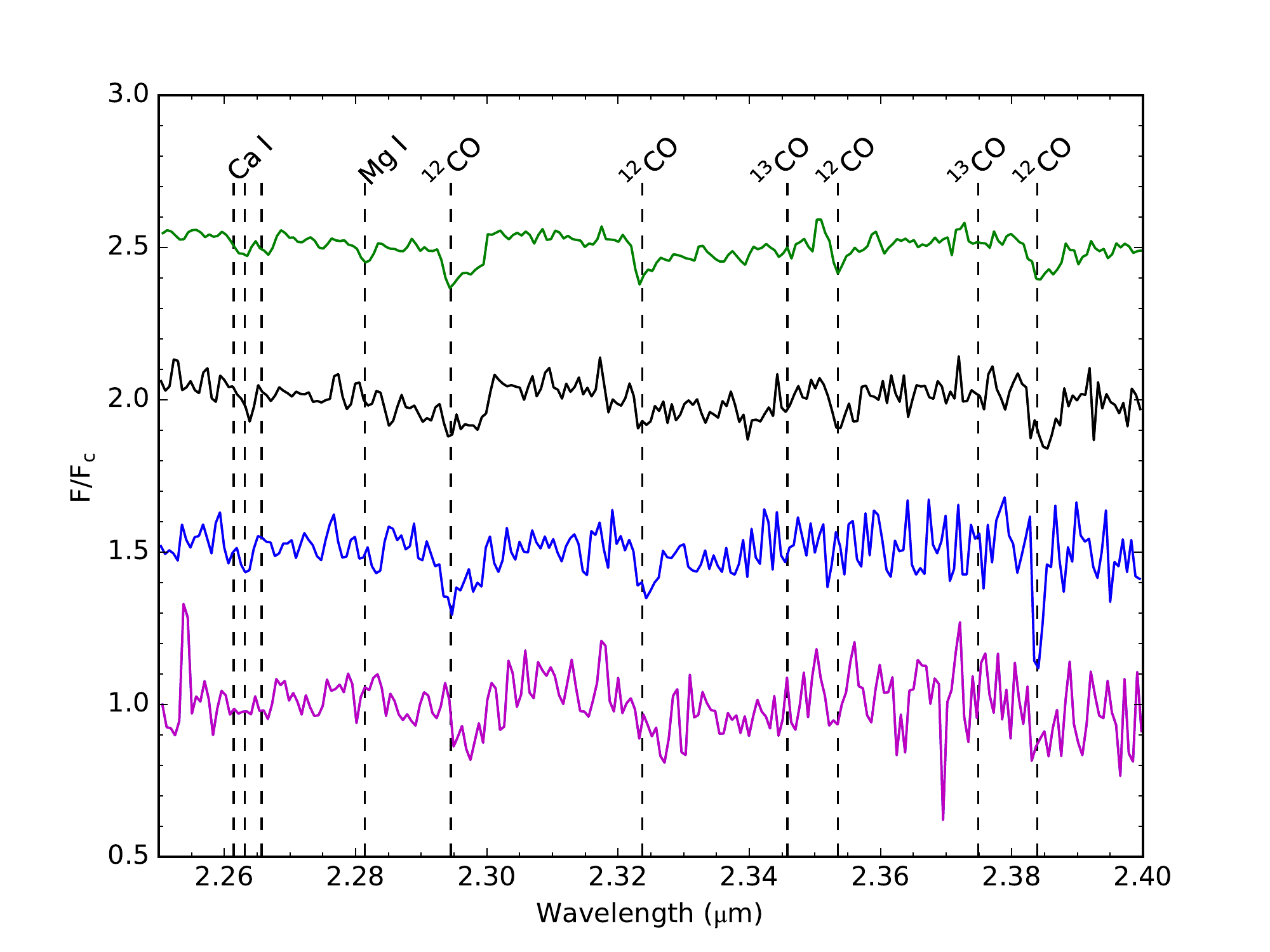}
\caption{Spectral features from the photosphere of star B (green) are still present in the intermediate (black) and faint phase spectra (blue, November; pink, December), which have been normalized and offset for comparison. The most prominent absorption lines are the CO bandheads, which are characteristic of K giants. There does not seem to be a clear relationship between the brightness of the star and the strength of the features.}
\label{COlines}
\end{figure}

Star B has a K1 spectral type, which was assigned when it re-emerged in 2012 \citep{Capelo2012}. The strongest features we detect from its photosphere are the CO band heads and atomic lines between $\sim$2.25 and 2.40 $\mu$m. Star A, which is a K6/K7 object, should have similar features at these wavelengths despite its slightly cooler temperature \citep{FS2000}. However, we expect that the contribution from star A is negligible in these data. The absorption lines from star B are most prominent in the bright phase spectrum (where the signal-to-noise ratio is highest), but stellar features can still be clearly identified in the other three spectra (see Figure \ref{COlines}). The equivalent width measurements from the bright spectrum can be used to verify the spectral classification of star B and constrain the amount of absorption we expect to see at the intermediate and faint phases. We can then remove the stellar contribution from these spectra. 

\begin{figure}
\centering
\includegraphics[width=1.0\linewidth]
{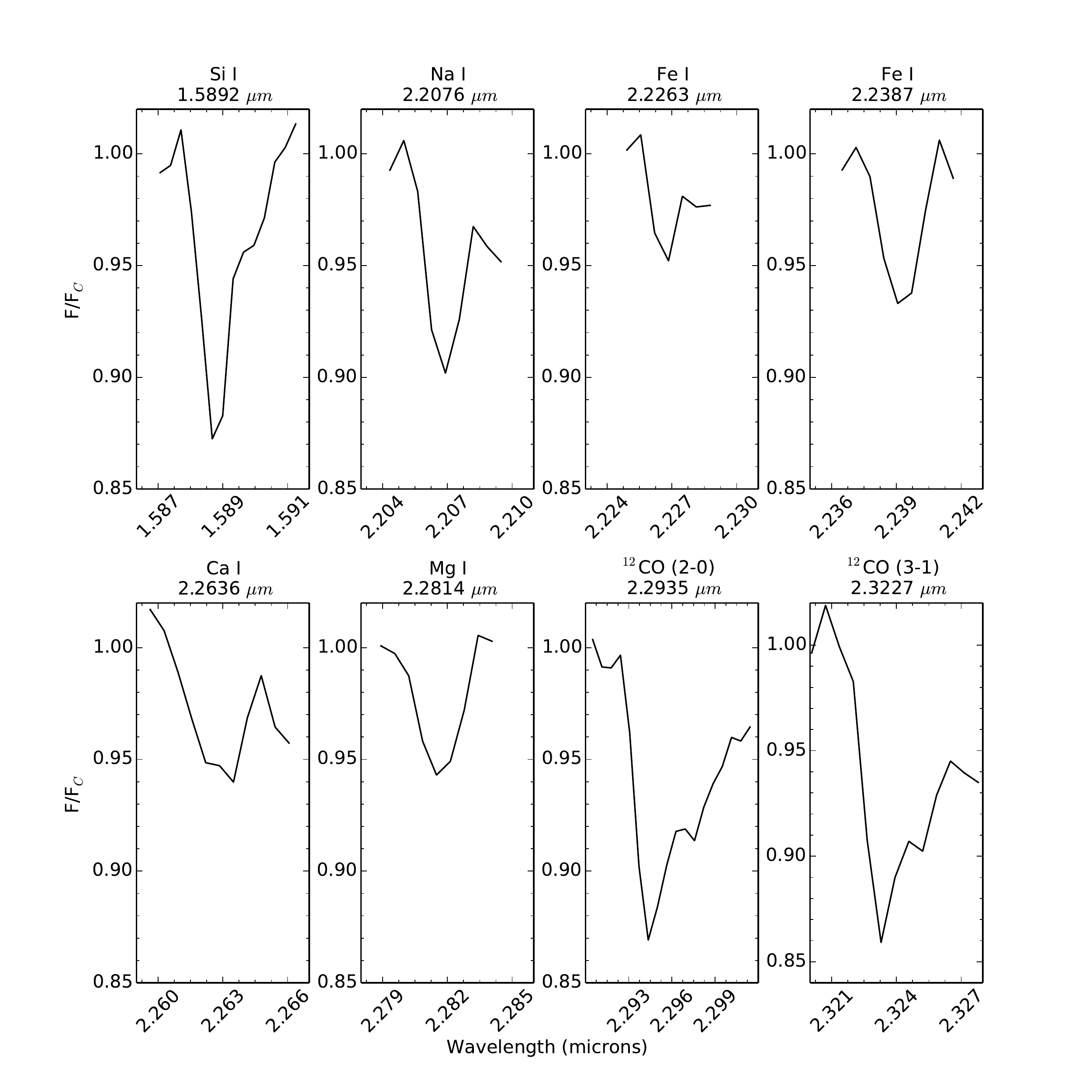}
\caption{The line strength ratios of the atomic absorption lines and CO band heads in the bright phase spectrum of KH 15D are consistent with a $\sim$4000-4500 K stellar photosphere. Star B also falls in between the dwarf and giant relationships outlined by the Na I, Ca I, and CO features, as expected for an object that is contracting to the main sequence.}
\label{brightEWfigure}
\end{figure}

\cite{FS2000} showed that absorption line strength ratios in the near-IR are correlated with stellar temperature for cool giants. In order to compare the photospheric absorptions of star B to those seen in other K giants, the equivalent widths of six atomic lines and two CO features were measured (see Figure \ref{brightEWfigure}, Table \ref{brightEWtable}). The line strength quantities with the strongest temperature dependences, as shown by \cite{FS2000}, are $EW_{CO \, (2.29)} / EW_{Mg \, I}$ and $EW_{Fe \, I \, (2.2263)} + EW_{Fe \, I \, (2.2387)}$. The ratios $\log{EW_{CO \, (1.62)} / EW_{Si \, I}}$ and $\log{EW_{CO \, (1.62)} / EW_{CO \, (2.29)}}$ also follow clear relationships with temperature, but the $^{12}$CO (6-3) line at 1.62 $\mu$m was not resolved in our data. 

Values of $EW_{CO \, (2.29)} / EW_{Mg \, I} = 5.7 \pm 0.1$ and $EW_{Fe \, I \, (2.2263)} + EW_{Fe \, I \, (2.2387)} = 2.31 \pm 0.03$ \AA \, were obtained from the bright phase spectrum. The measurements indicate a temperature of $\sim$4000-4500 K, which is consistent with the effective temperature of a $\sim$K1-K3 giant \citep{VB1999}. In addition, the quantity $EW_{Na \, I} + EW_{Ca \, I} = 4.47 \pm 0.06$ \AA \, and the equivalent width of the $^{12}$CO (2-0) line at 2.29 $\mu$m ($EW = 6.8 \pm 0.1$ \AA) place KH 15D in between the giant and dwarf branches described by \cite{IT2004}, as expected for a WTTS that is contracting to the main sequence. Our data, therefore, confirm the spectral class of star B as around K1, and indicate that it is a sub-giant. 

\begin{deluxetable}{ccc}
\tablewidth{0 pt}
\tabletypesize{\scriptsize}
\tablecaption{Equivalent Widths of Bright Phase Features \label{brightEWtable}}
\tablehead{
\colhead{Feature} & \colhead{Wavelength ($\mu$m)} & \colhead{EW (\AA)} 
}
\startdata
Si I & 1.5892 & 2.44 $\pm$ 0.08 \\
Na I & 2.2076 & 2.51 $\pm$ 0.05 \\
Fe I & 2.2263 & 0.89 $\pm$ 0.01 \\
Fe I & 2.2387 & 1.42 $\pm$ 0.02 \\
Ca I & 2.2636 & 1.96 $\pm$ 0.03 \\
Mg I & 2.2814 & 1.18 $\pm$ 0.02 \\
$^{12}$CO (2-0) & 2.2935 & 6.8 $\pm$ 0.1 \\
$^{12}$CO (3-1) & 2.3227 & 5.1 $\pm$ 0.1 \\
\enddata
\end{deluxetable} 

\subsection{Magnetosphere of Star B}

\begin{figure}
\centering
\includegraphics[width=0.8\linewidth]
{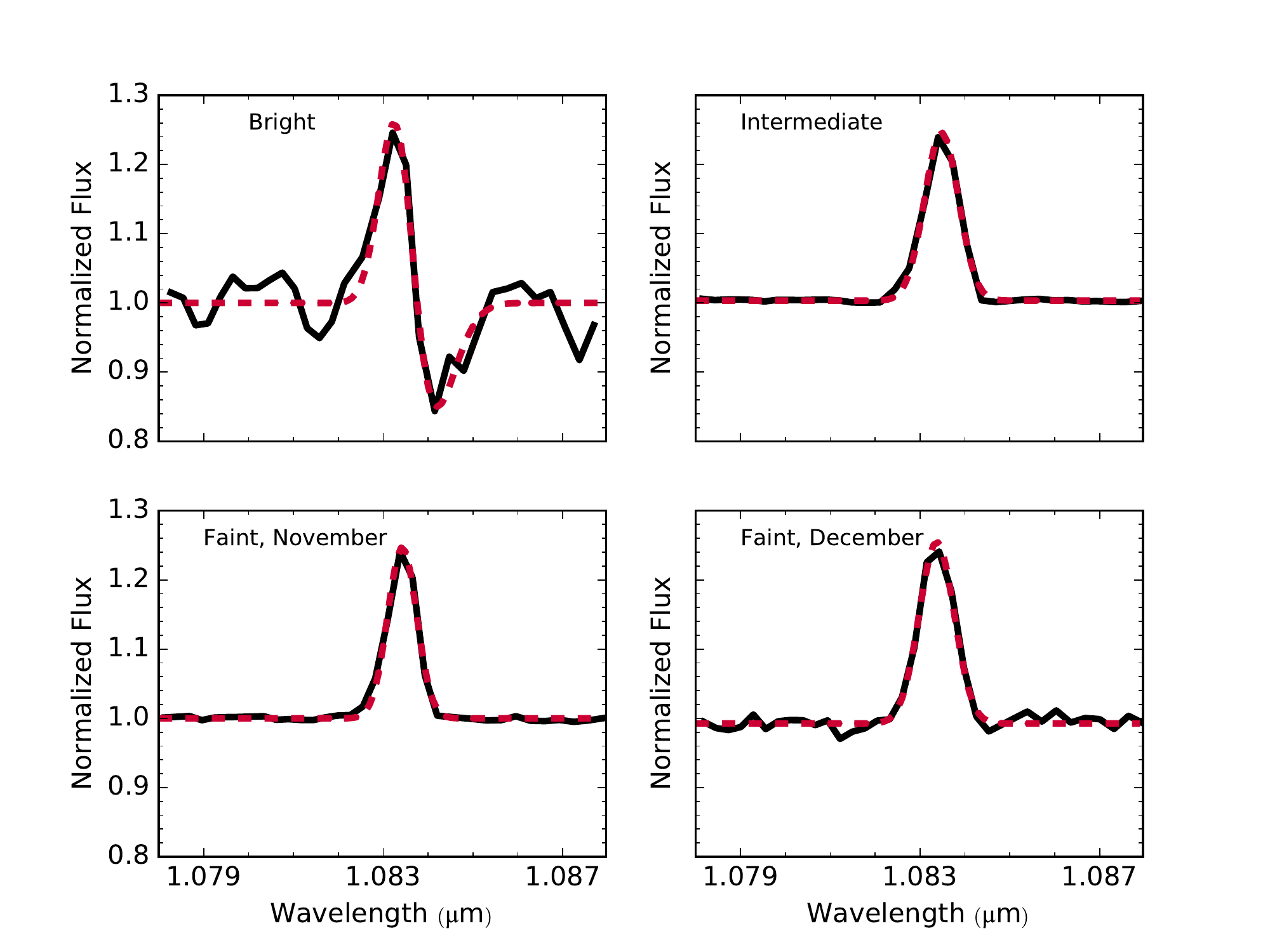}
\caption{Gaussian fits (red, dashed lines) to the He I $\lambda$10830 emission seen in all four spectra (black, solid lines). A red-shifted absorption component was also detected in the bright spectrum, indicating that material is accreting onto the stars. The normalized emission lines from the other three spectra have been scaled down to the level of the bright phase feature, showing that the absorption was not present at the same level during the intermediate and faint phases.}
\label{Helines}
\end{figure}

\begin{figure}
\centering
\includegraphics[width=0.8\linewidth]
{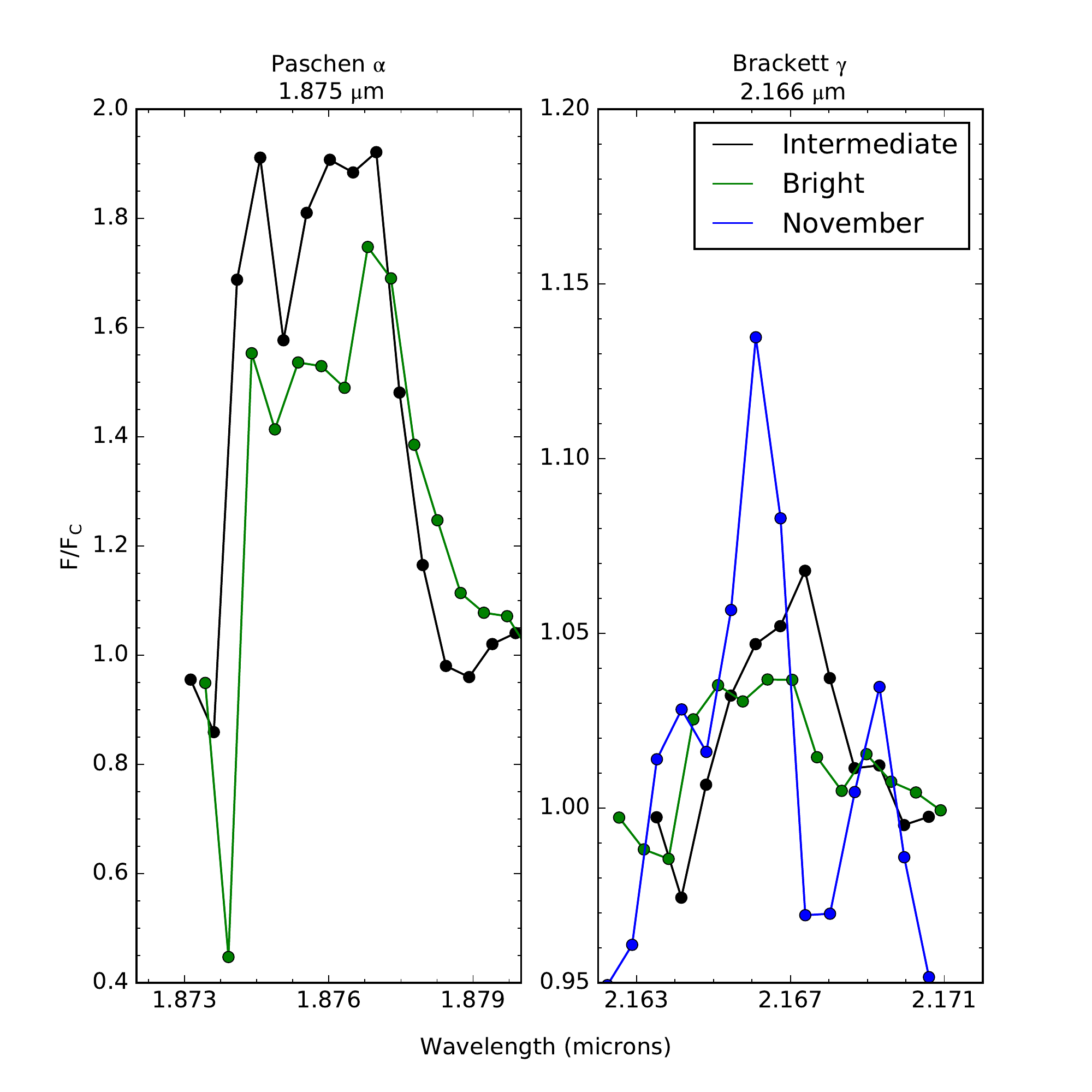}
\caption{The Paschen $\alpha$ and Brackett $\gamma$ emission lines are present in the bright and intermediate phase spectra, but only Brackett $\gamma$ is observed in the November faint spectrum. Neither line was detected in December. Both features have been divided by the continuum flux in this figure to make it easier to compare line strengths between phases.}
\label{neutralHlines}
\end{figure}

In addition to photospheric absorption, star B shows He I $\lambda$10830 emission at all three orbital phases and Paschen $\alpha$ and Brackett $\gamma$ lines in the bright and intermediate phase spectra. Equivalent widths and absolute line fluxes were measured, and the results are presented in Table \ref{Hetable} (He I) and Table \ref{neutralH} (Paschen $\alpha$ and Brackett $\gamma$). Figure \ref{Helines} shows the He I line profiles from all four spectra, and the hydrogen lines are plotted in Figure \ref{neutralHlines}. 

\begin{deluxetable}{lcc}
\tablewidth{0 pt}
\tabletypesize{\footnotesize}
\tablecolumns{5}
\tablecaption{Equivalent Width Measurements for He I $\lambda$10830 Emission \label{Hetable}}
\tablehead{
\colhead{Phase} & \colhead{Equivalent Width} \vspace{0.05cm} & \colhead{Absolute Line Flux} \vspace{0.05cm}  \\
\colhead{} & \colhead{ \AA} & \colhead{($10^{-18}$ W m$^{-2}$)}
}
\startdata
Bright (Emission Only) & $-2.10 \pm 0.02$ & $2 \pm 1$  \\
Bright (Emission + Absorption) & $-2.21 \pm 0.02$ & $3 \pm 20$ \\
 & $1.47 \pm 0.01$ & $2 \pm 20$ \\
Intermediate & $-26.2 \pm 0.2$ & $3.0 \pm 0.2$ \\
Faint, November & $-44.1 \pm 0.9$ & $2.2 \pm 0.2$ \\
Faint, December & $-30.5 \pm 0.6$ & $1.4 \pm 0.2$ \\
\
\enddata
\end{deluxetable} 

\begin{deluxetable}{cccc}
\tablewidth{0 pt}
\tabletypesize{\footnotesize}
\tablecolumns{5}
\tablecaption{Equivalent Width Measurements for Atomic Hydrogen Emission \label{neutralH}}
\tablehead{
\colhead{Phase} & \colhead{Feature} \vspace{0.05cm} & \colhead{Equivalent Width} \vspace{0.05cm} & \colhead{Absolute Line Flux} \vspace{0.05cm} \\
\colhead{} & \colhead{$\left( \mu m \right)$} & \colhead{\AA} & \colhead{($10^{-18}$ W m$^{-2}$)}
}
\startdata
Bright & Paschen $\alpha$ & $-20 \pm 1$ & $7 \pm 4$ \\
& Brackett $\gamma$ & $-1.17 \pm 0.03$ & $0.3 \pm 0.1$ \\
Intermediate & Paschen $\alpha$ & $-30 \pm 1$ & $1.3 \pm 0.3$ \\
& Brackett $\gamma$ & $-1.49 \pm 0.05$ & $0.038 \pm 0.009$ \\
Faint, November & Brackett $\gamma$ & $-1.02 \pm 0.03$ & $0.022 \pm 0.007$ \\
\enddata
\end{deluxetable} 

A red-shifted absorption component is observed in the He I feature at the bright phase, which is indicative of gas falling in toward the stars. This inverse P-Cygni shape is additional confirmation that KH 15D is still weakly accreting, although the spectral resolution of our data is not high enough to apply a detailed model of the accretion flow, as has been done with other CTTS (e.g. \citet{Fischer2008}). A two-component fit to the bright phase feature gave equivalent widths of $-2.21 \pm 0.02$ \AA \, for the emission and $1.47 \pm 0.01$ \AA \, for the absorption, while a single component fit gave an equivalent width of $-2.10 \pm 0.02$ \AA \, for emission only. However, the high uncertainty in the corresponding line fluxes makes it difficult to determine whether the absorption is significant.   

The He I emission profile has its highest total flux in the intermediate spectrum and its lowest in the December faint phase spectrum, indicating variable emission with orbital phase. Furthermore, the line strengths between the two faint spectra are not consistent with each other. It is possible that a heating event is responsible for the increased emission in the intermediate spectrum and/or the November faint spectrum. \cite{Zeng2014} studied enhanced He emission at UV and near-infrared wavelengths following a C-class solar flare and attributed the increase in line flux to photoionization followed by recombination during the flaring event. YSOs are highly variable objects that can experience frequent flaring events, which would increase the strength of He I emission for a brief time afterwards. 

It is difficult to determine with certainty whether the varying He I emission strengths in our data can be attributed to flares, but any heating events that occur could be expected to cause an increase in the strengths of other features as well. The hydrogen Paschen $\alpha$ and Brackett $\gamma$ features are observed in the bright and intermediate phase spectra, implying that they should at least be detectable in the November faint spectrum. However, only one of the two lines that were measured in the bright and intermediate spectra is distinguishable in the November spectrum, and neither line was present in the data from December. The heating event that caused the strengthening of the He I feature may have had either no impact or an inverse impact on the atomic hydrogen in the stellar magnetosphere. Perhaps the temperature increase implied by the He I outburst was sufficient to increase the degree of ionization of the magnetospheric hydrogen and actually reduce the amount of emission in the H I features. We leave further examination of this interesting possible anti-correlation to future work.

\subsection{Inner Jet}

The prominent H$_2$ ro-vibrational emission features appearing in the intermediate and faint phase spectra have previously been attributed to the bipolar outflow associated with the system's jet \citep{Tokunaga2004, Deming2004}. Table \ref{transitions} lists the 10 H$_2$ lines that were identified in the intermediate and faint phase spectra, with the five features measured by \cite{Deming2004} denoted by asterisks. Equivalent widths were measured for all 10 lines, although values for several lines that were noisy in the intermediate and December faint phase spectra have been omitted. The continuum emission, $F_c$, was assumed to be flat in the immediate vicinity of each line and was estimated by fitting a constant to the region. 

\begin{deluxetable}{ccccc}
\tablewidth{0 pt}
\tabletypesize{\scriptsize}
\tablecaption{Near-Infrared H$_2$ Transitions \citep{Turner1977, Dabrowski1984} \label{transitions}}
\tablehead{
\colhead{Central Wavelength} \vspace{0.05cm} & \colhead{Transition} & \colhead{Energy} \vspace{0.05cm} & $g_j$ & \colhead{$A_{ij}$} \vspace{0.05cm} \\
\colhead{($\mu$m)} & \colhead{} & \colhead{(K)} & \colhead{} & \colhead{(10$^{-7}$ s$^{-1}$)} \\
}
\startdata
2.0338 & S2 (1-0) & 7584 & 9 & 3.98 \\
2.1218* & S1 (1-0) & 6956 & 21 & 3.47 \\
2.2235* & S0 (1-0) & 6471 & 5 & 2.53 \\
2.2477* & S1 (2-1) & 12550 & 21 & 4.98 \\
2.4066* & Q1 (1-0) & 6149 & 9 & 4.29 \\
2.4134 & Q2 (1-0) & 6471 & 5 & 3.03 \\
2.4237* & Q3 (1-0) & 6956 & 21 & 2.78 \\
2.4375 & Q4 (1-0) & 7586 & 9 & 2.65 \\
2.4548 & Q5 (1-0) & 8365 & 33 & 2.55 \\
2.4756 & Q6 (1-0) & 9286 & 13 & 2.45 \\
\enddata
\tablenotetext{*}{Features also detected by \citep{Deming2004}}
\end{deluxetable}  

\begin{figure}
\centering
\includegraphics[width=0.9\linewidth]
{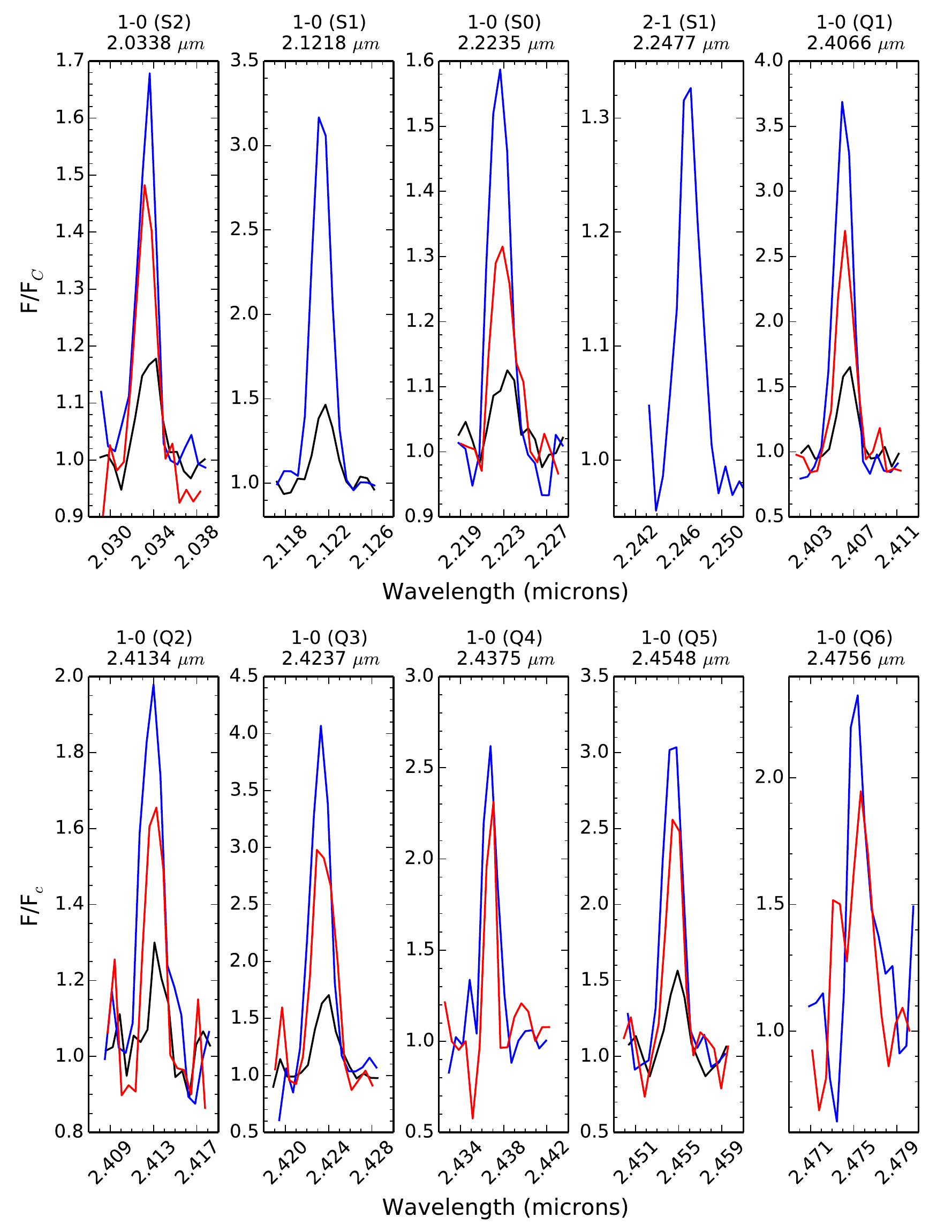}
\caption{The equivalent widths of the H$_2$ lines appear to increase from the intermediate phase (black) to the faint phase (November in blue, December in red) spectra. Cloudy weather on the night of the November observing run resulted in a less accurate flux calibration for that spectrum than the others, but the absolute line fluxes are about the same in the intermediate and December faint phase spectra (see Table \ref{absflux}). This is what we would expect for lines that originate in a region far from the star.}
\label{H2lines}
\end{figure}

Figure \ref{H2lines} shows that the contrast between the H$_2$ lines and the continuum appears to increase as KH 15D becomes fainter, in accordance with expectation if the emission lines are arising from a source well away from the star itself. Although the November and December faint spectra were obtained at similar phases, the December lines consistently have smaller equivalent widths (see Table \ref{absflux}). The wavelengths of the H$_2$ emission features are contained within the \emph{K} band, so broadband photometry from the two nights that the spectra were collected could help to account for the deviation in line strengths between cycles. However, the \emph{K} band photometry from the December observing night was highly uncertain and can't be used as a reliable measurement. 

The absolute fluxes for each H$_2$ line in the three spectra are presented in Table \ref{absflux} and show that, on average, the line fluxes are more consistent between the intermediate and December spectra than between the November and December spectra. As discussed previously, cloudy weather during the November observing run resulted in a poor flux calibration for that spectrum. It is also possible that poor seeing conditions induced by the cloudiness may have scattered more light from the extended jet region into the slit in November than in the December or intermediate spectrum. This could lead to an apparent increased flux in the line. In any event, the uncertainties introduced by the flux calibration or variable extent of the region probed are minimized when the line flux ratios of the H$_2$ features are analyzed, as is the case here.  

\begin{deluxetable}{ccccccc}
\tablewidth{0 pt}
\tabletypesize{\scriptsize}
\tablecaption{Near-Infrared H$_2$ Equivalent Widths and Absolute Fluxes \label{absflux}}
\tablehead{
\colhead{Central Wavelength} \vspace{0.05cm} & \colhead{EW$_{I}$ \tablenotemark{a}} \vspace{0.05cm} & \colhead{F$_I$} \vspace{0.05cm} & \colhead{EW$_{F, N}$\tablenotemark{b}} \vspace{0.05cm} & \colhead{F$_{F, N}$} \vspace{0.05cm} & \colhead{EW$_{F, D}$\tablenotemark{c}} \vspace{0.05cm} & \colhead{F$_{F, D}$} \vspace{0.05cm} \\
\colhead{($\mu$m)} & \colhead{ (\AA)} & \colhead{(10$^{-19}$ W m$^{-2}$)} & \colhead{ (\AA)} & \colhead{(10$^{-19}$ W m$^{-2}$)} & \colhead{ (\AA)} & \colhead{(10$^{-19}$ W m$^{-2}$)}
}
\startdata
2.0338 & $-3.64 \pm 0.03$ & $0.9 \pm 0.2$ & $-14.5 \pm 0.2$ & $2.0 \pm 0.2$ & $-7.7 \pm 0.2$ & $1.0 \pm 0.3$ \\
2.1218 & $-9.14 \pm 0.07$ & $2.0 \pm 0.2$ & $-48.1 \pm 0.5$ & $5.9 \pm 0.1$ & - & - \\ 
2.2235 & $-3.80 \pm 0.03$ & $0.7 \pm 0.2$ & $-12.3 \pm 0.1$ & $1.2 \pm 0.1$ & $-7.92 \pm 0.08$ & $0.72 \pm 0.08$ \\ 
2.2477 & - & - & $-6.7 \pm 0.1$ & $0.6 \pm 0.1$ & - & - \\ 
2.4066 & $-10.8 \pm 0.2$ & $1.6 \pm 0.3$ & $-45 \pm 3$ & $4.6 \pm 0.6$ & $-27 \pm 2$ & $2.3 \pm 0.5$ \\ 
2.4134 & $-5.4 \pm 0.2$ & $0.7 \pm 0.5$ & $-24 \pm 1$ & $2.1 \pm 0.3$ & $-12.4 \pm 0.7$ & $1.0 \pm 0.4$ \\ 
2.4237 & $-16.1 \pm 0.2$ & $2.2 \pm 0.2$ & $-64 \pm 2$ & $5.4 \pm 0.3$ & $-51 \pm 2$ & $4.0 \pm 0.3$ \\
2.4375 & - & - & $-26.4 \pm 0.5$ & $2.2 \pm 0.2$ & $-16.6 \pm 0.6$ & $1.3 \pm 0.3$ \\ 
2.4548 & $-9.7 \pm 0.2$ & $1.3 \pm 0.2$ & $-46 \pm 2$ & $3.5 \pm 0.3$ & $-31 \pm 1$ & $2.4 \pm 0.3$ \\
2.4756 & - & - & $-32 \pm 3$ & $2.1 \pm 0.7$ & $-22 \pm 3$ & $1.4 \pm 0.7$ \\  
\enddata
\tablenotetext{a}{$I$: intermediate phase}
\tablenotetext{b}{$F, N$: faint phase, November}
\tablenotetext{c}{$F, D$: faint phase, December}
\end{deluxetable}  

\cite{Deming2004} attributed the H$_2$ features to shock heating of ambient hydrogen within a thermally excited bipolar outflow, based on the measured line intensity ratio $2-1 \, S(1) / 1-0 \, S(1) = 0.24 \pm 0.05$. This result is well below the value of $\sim$0.5 expected for fluorescent lines. \cite{Deming2004} measured the intensity of five of the hydrogen lines in a spectrum acquired at phase $\sim$0.17 (similar to our intermediate phase) and found a best-fit temperature of $T \sim 2800 \pm 300$ K for the emitting region. However, it was noted that the observed line strength ratios did not agree with the expected values for this temperature within the measured uncertainties. 

Table \ref{fluxratiotable} gives the line flux ratios of the H$_2$ features in the most recent spectra (calculated from the absolute line fluxes in Table \ref{absflux}). The intensity of the (1-0) Q1 line $\left( \lambda = 2.4066 \,  \mu \text{m} \right)$ was used as the normalizing flux instead of the (1-0) S1 line $\left( \lambda = 2.1218 \,  \mu \text{m} \right)$, which had a low S/N in the December spectrum. Although the equivalent width of the (1-0) Q3 line was greater than that of the 1-0 (Q1) line in all three spectra, the uncertainties were lower for 1-0 (Q1). The (2-1) S1 feature $\left( \lambda = 2.477 \,  \mu \text{m} \right)$ could only be measured for the November spectrum. We obtained a ratio of 2-1 S(1)/1-0 S(1)$ = 0.10 \pm 0.02$, which again confirms that the source of the emission is primarily shock excitation. 

\begin{deluxetable}{ccccc}
\tablewidth{0 pt}
\tabletypesize{\scriptsize}
\tablecaption{Near-Infrared H$_2$ Line Flux Ratios (x / (1-0) Q1) \label{fluxratiotable}}
\tablehead{
\colhead{Central Wavelength} & \colhead{Transition} & \colhead{Intermediate} & \colhead{Faint} & \colhead{Faint } \\ 
\colhead{($\mu$m)}  &  & &  \colhead{(Nov. 2013)} & \colhead{(Dec. 2013)} \\
}
\startdata
2.0338 & 1-0 (S2) & $0.6 \pm 0.2$ & $0.43 \pm 0.08$ & $0.4 \pm 0.2$ \\
2.1218 & 1-0 (S1) & $1.3 \pm 0.2$ & $1.3 \pm 0.2$ & - \\
2.2235 & 1-0 (S0) & $0.5 \pm 0.1$ & $0.26 \pm 0.04$ & $0.31 \pm 0.08$ \\
2.2477 & 2-1 (S1) & - & $0.13 \pm 0.04$ & - \\
2.4066 & 1-0 (Q1) & 1.0 & 1.0 & 1.0 \\
2.4134 & 1-0 (Q2) & $0.5 \pm 0.3$ & $0.4 \pm 0.1$ & $0.4 \pm 0.2$ \\
2.4237 & 1-0 (Q3) & $1.4 \pm 0.3$ & $1.2 \pm 0.2$ & $1.7 \pm 0.4$ \\
2.4375 & 1-0 (Q4) & - & $0.3 \pm 0.1$ & $0.5 \pm 0.2$ \\
2.4548 & 1-0 (Q5) & $0.9 \pm 0.2$ & $0.8 \pm 0.1$ & $1.0 \pm 0.3$ \\
2.4756 & 1-0 (Q6) & - & $0.5 \pm 0.2$ & $0.6 \pm 0.3$ \\
\enddata
\end{deluxetable} 

\cite{Deming2004} compared the spectrum obtained at phase $\sim$0.17 to a spectrum taken out of eclipse (similar to our bright phase) in which the 1-0 S(1) feature was resolved. The strength of the feature in the bright spectrum was much weaker than it was when the star was fainter, and the difference in strength was attributed to the increase in continuum flux when the stellar surface was revealed. If this were true in our spectra, we would expect the line flux ratios to remain roughly constant with orbital phase, despite the difference in absolute flux. The values in Table \ref{fluxratiotable} for each line do not always agree within the measured uncertainties for all three spectra. However, we do not know the exact location of the emitting region detected in each of the three spectra and could simply be probing slightly different areas.  

A thermal excitation temperature $\left( T \right)$ can be determined by assuming $\ln \left( \frac{F}{g A} \right) = -E / T$, with the properties of the (1-0) Q1 line used for reference. Table \ref{transitions} lists the line parameters $A$, $g$, and $E$ for each of the ten H$_2$ lines we observed. The most reliable flux ratios from Table \ref{fluxratiotable} were used to determine the temperature of the emitting region. The resulting temperatures for the shock-excited H$_2$ are $2400 \pm 300$ K (November), $3000 \pm 200$ K (December), and $2800 \pm 500$ K (intermediate). The value obtained from the December spectrum is likely more uncertain than $\pm 200$ K, given the low S/N of some of the features. The temperatures from the November and intermediate spectra are consistent with each other and the $2800 \pm 300$ K result given by \cite{Deming2004}. At the time the \cite{Deming2004} observations were conducted, the entire surface of star A was directly visible near apastron, and star B was fully occulted at all phases. When our spectra were acquired in 2013, star A was fully occulted at all phases and only part of star B was directly visible near apastron. The geometry of the system was quite different when the two sets of spectra were obtained, so the consistency between temperatures measured during two different epochs further confirms that the emission comes from the base of the bipolar outflow. 

\begin{figure}
\centering
\includegraphics[width=0.9\linewidth]
{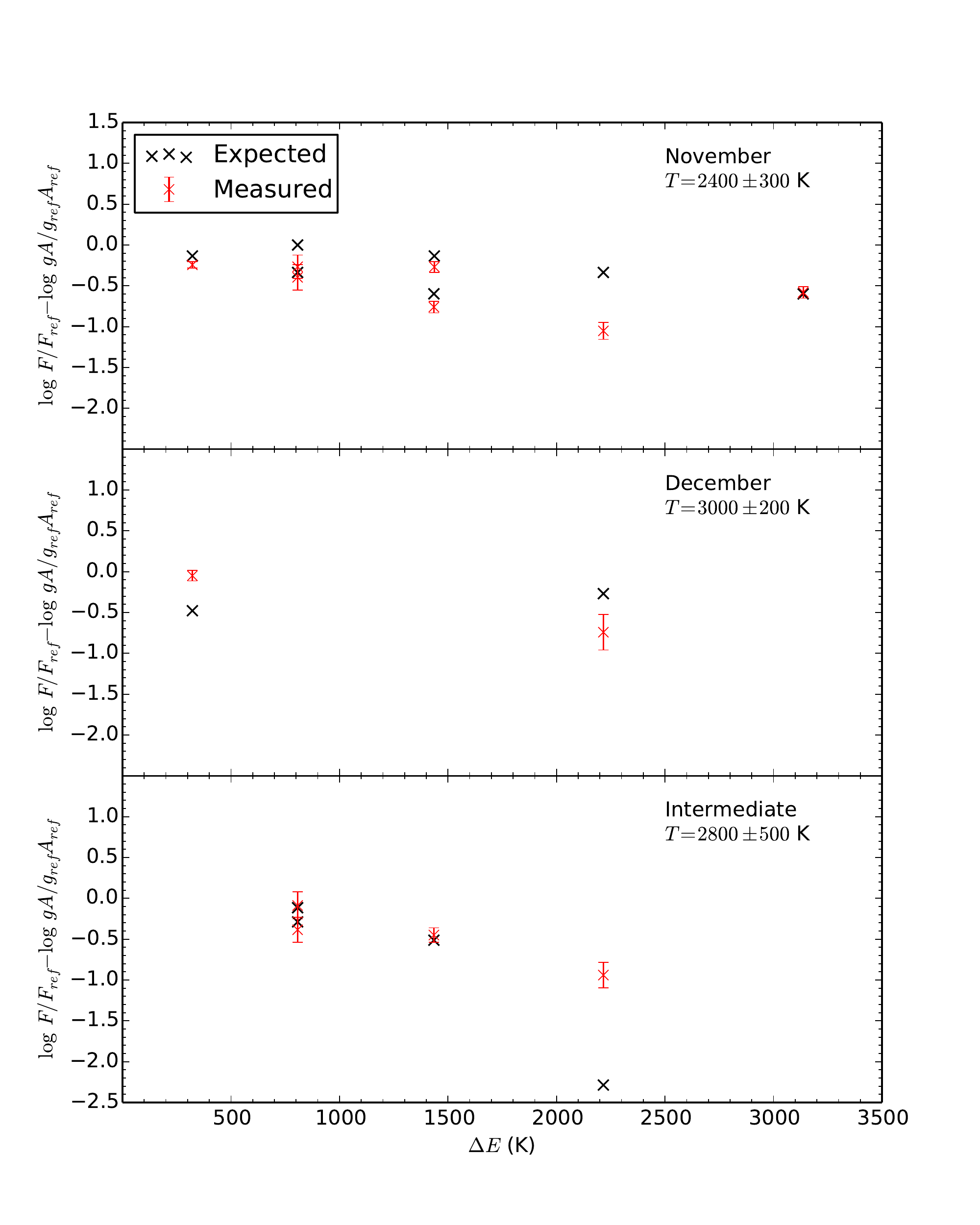}
\caption{A comparison of the measured flux ratios to those expected for shock-excited H$_2$ at the best-fit temperatures derived in this section shows that our measured values are consistent with each other and with the value of $2800 \pm 300$ K obtained by \cite{Deming2004}. The agreement between our data and previous work done when star A was the only directly visible component further confirms that the features originate in the bipolar outflow detected by \cite{Tokunaga2004}.}
\label{tempcomp}
\end{figure}

Figure \ref{tempcomp} compares the measured flux ratios to the expected values for H$_2$ at the best-fit temperatures. Lines with $\frac{\sigma_{F_{line} / F_{ref}}}{F_{line} / F_{ref}} \geq 0.20$ have been omitted. The December spectrum only had two features that fell under this limit, again implying that the uncertainty in the temperature obtained from these data is actually greater than for the other two spectra. Although the measured flux ratios do not agree with the best-fit temperature for each individual feature, there is strong agreement between the values we have obtained from the spectra and what is expected for shock-excited H$_2$ in a bipolar outflow. 

To summarize this discussion, we find no convincing evidence for variations in the flux of the H$_2$ lines either on the orbital period or over the longer time scale represented by the difference in epochs between these data and the original studies. The current observations were not designed to study the jet emission, which is quite extended and should be investigated with a long slit aligned along the jet. Here the exact region probed by the H$_2$ lines is less clear, but it is apparent that it does not suffer variable obscuration by the ring. The line ratio data support the previous interpretation that it is shock heated ambient gas. 

\subsection{Reflectance Properties of the Circumbinary Ring}

The scattered light from KH 15D reaches a maximum during the intermediate phase \citep{Agol2004, Silvia2008, A16}, when direct starlight has declined but the bulk of the optically thick ring material is not yet between us and the star. To check whether the scattered light carries spectral signatures of the dust, reflectance spectra were created from the GNIRS data based on the procedure of \cite{Vilas1984}, who removed the solar contribution from spectra of the surface of Mercury by dividing out a standard spectrum of the Sun. \citet{Debes2013} applied a similar procedure to near-IR spectra of TW Hya. Analogously, the intermediate phase spectrum of KH 15D was divided by the bright phase spectrum, where the features of star B are dominant. The most prominent emission features, such as He I from the magnetosphere and H$_2$ from the inner jet, were removed from the spectrum, which was normalized to the flux level at 0.8 $\mu$m. A 50 point moving average was applied to the data to increase the S/N of the remaining broad continuum while preserving the resolution of the original spectrum.  

\begin{figure}
\centering
\includegraphics[width=0.9\linewidth]
{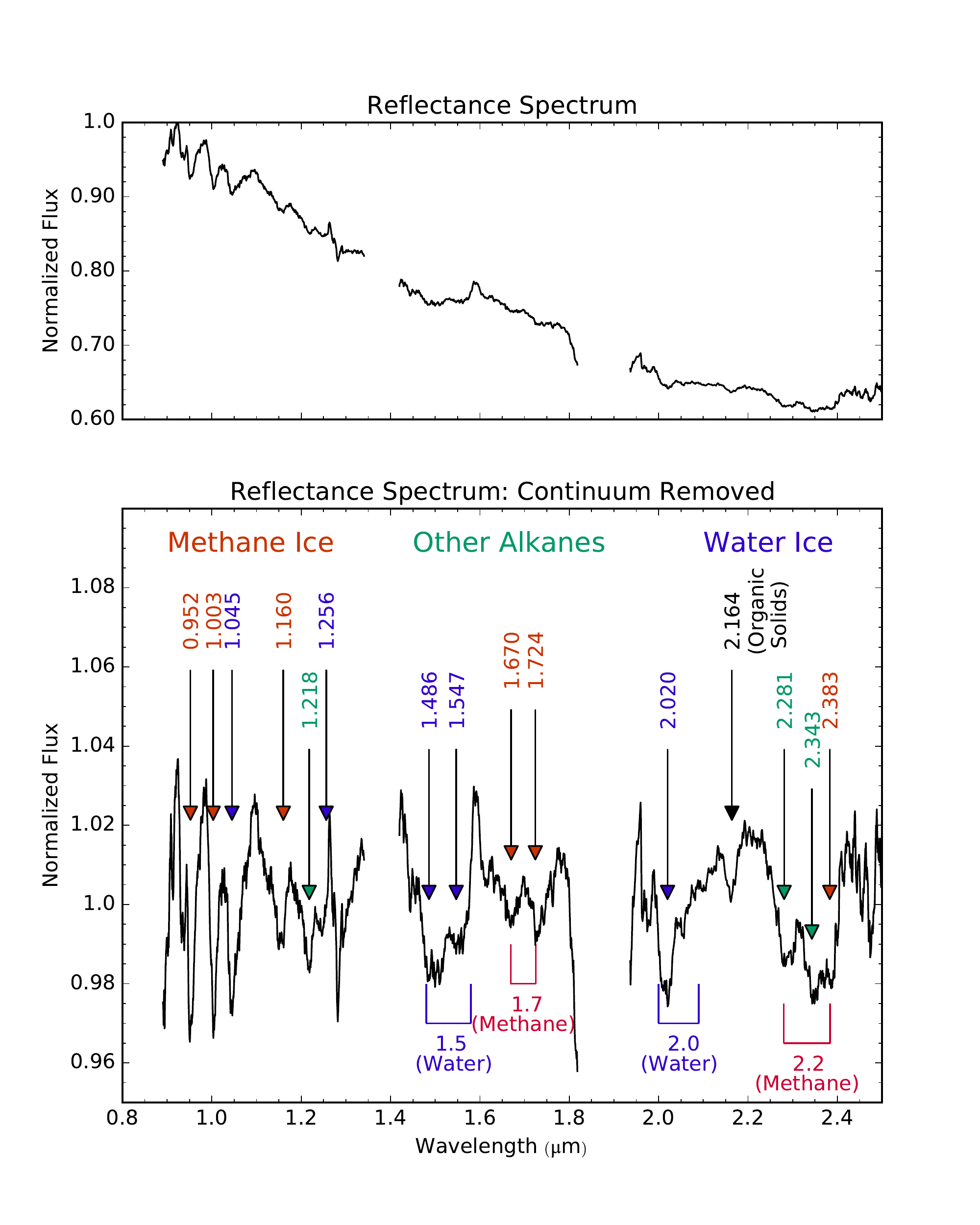}
\caption{The reflectance spectrum of KH 15D (top panel, obtained by dividing the bright phase spectrum from the intermediate phase data) shows a steep blue slope, which is indicative of both forward scattering and reflectance of water ice. We have removed the contribution from scattered light (bottom panel) and marked reflectance features that are indicative of water ice (blue; \citet{Clark1980}), methane ice (red; \citet{QS1997}, \citet{Clark2009}, \citet{Grundy2002}), other alkanes (green; \citet{QS1997}, \citet{Clark2009}), and organic solids (black; \citet{Cloutis1989}). All the features are listed in Table \ref{reflectancefeatures}.}
\label{reflectancespectrum}
\end{figure}

The slope of the resulting reflectance spectrum decreases towards longer wavelengths (see Figure \ref{reflectancespectrum}, top panel), which is consistent with the forward scattering profile that is expected when the star is at the edge of the ring but not yet completely occulted \citep{A16}. Forward scattering at shorter wavelengths peaks at a higher relative flux level than the contribution from longer wavelengths near the ring edge, which causes the system to become bluer at this phase. The spectral slope is also indicative of reflectance by water ice. It is not as steep as the spectra of many icy bodies in the solar system, which are typically darkened and reddened by the presence of irradiated and/or exogenic organic, non-icy compounds (\citet{Cruikshank2005} and references therein). The spectrum is more consistent with small ice grains. \citet{Filacchione2012} found that the slope of modeled reflectance spectra of pure water ice becomes shallower with decreasing grain size. The slope of the KH 15D spectrum appears to be more similar to that of the near-infrared spectra of Comet 17P/Holmes, whose features were attributed to a population of micron-sized grains in the coma \citep{Yang2009}. 

\begin{figure}
\centering
\includegraphics[width=0.9\linewidth]
{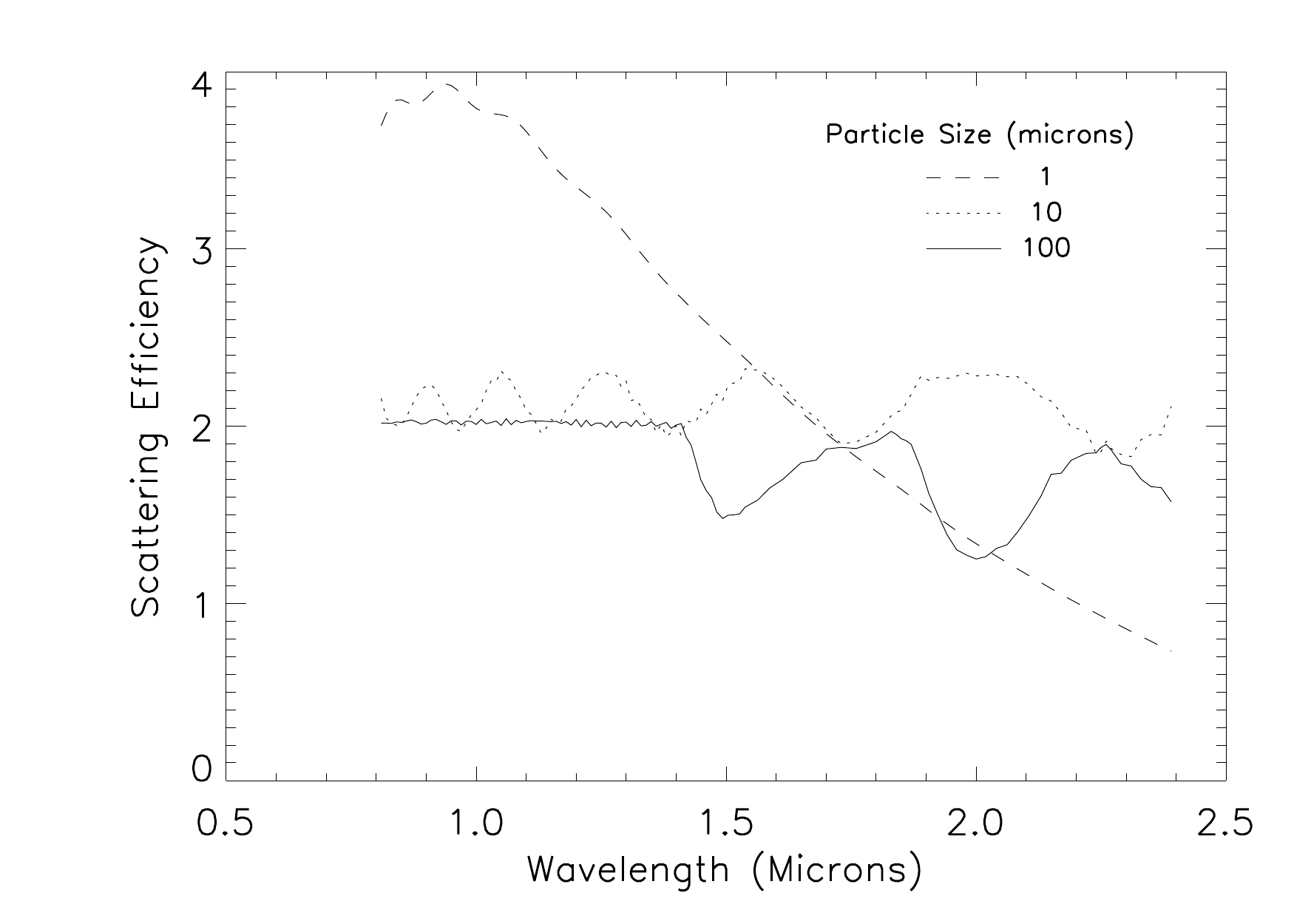}
\caption{Mie scattering models for various sizes of spherical ice grains. The slope of the KH 15D spectrum indicates that 1 $\mu$m grains are responsible for the scattered light, but larger particles are required to produce the observed water ice absorption features at 1.5 and 2.0 $\mu$m.}
\label{miescatteringsizes}
\end{figure}

To further test the hypothesis that small water ice grains are a primary component of the reflecting material in the disk, and to constrain grain sizes, we have performed Mie scattering calculations for spherical ice grains with a variety of sizes. We used the mie\_single code (McGarragh 2015; URL: http://eodg.atm.ox.ac.uk/MIE/) with optical constants for ice from \citet{Warren1984}. In Figure \ref{miescatteringsizes}, we show the scattering efficiency of particles of different sizes over the relevant wavelength range. Grains of $\sim$1 $\mu$m in size scatter very efficiently in the blue end of the range, while larger particles have a flatter spectral response and begin to show the characteristic absorption bands at 1.5 and 2.0 $\mu$m. We require a mixture of particle sizes, including some as small as 1 $\mu$m and some much larger, to account for the observed blue slope and absorption features.

\begin{figure}
\centering
\includegraphics[width=0.9\linewidth]
{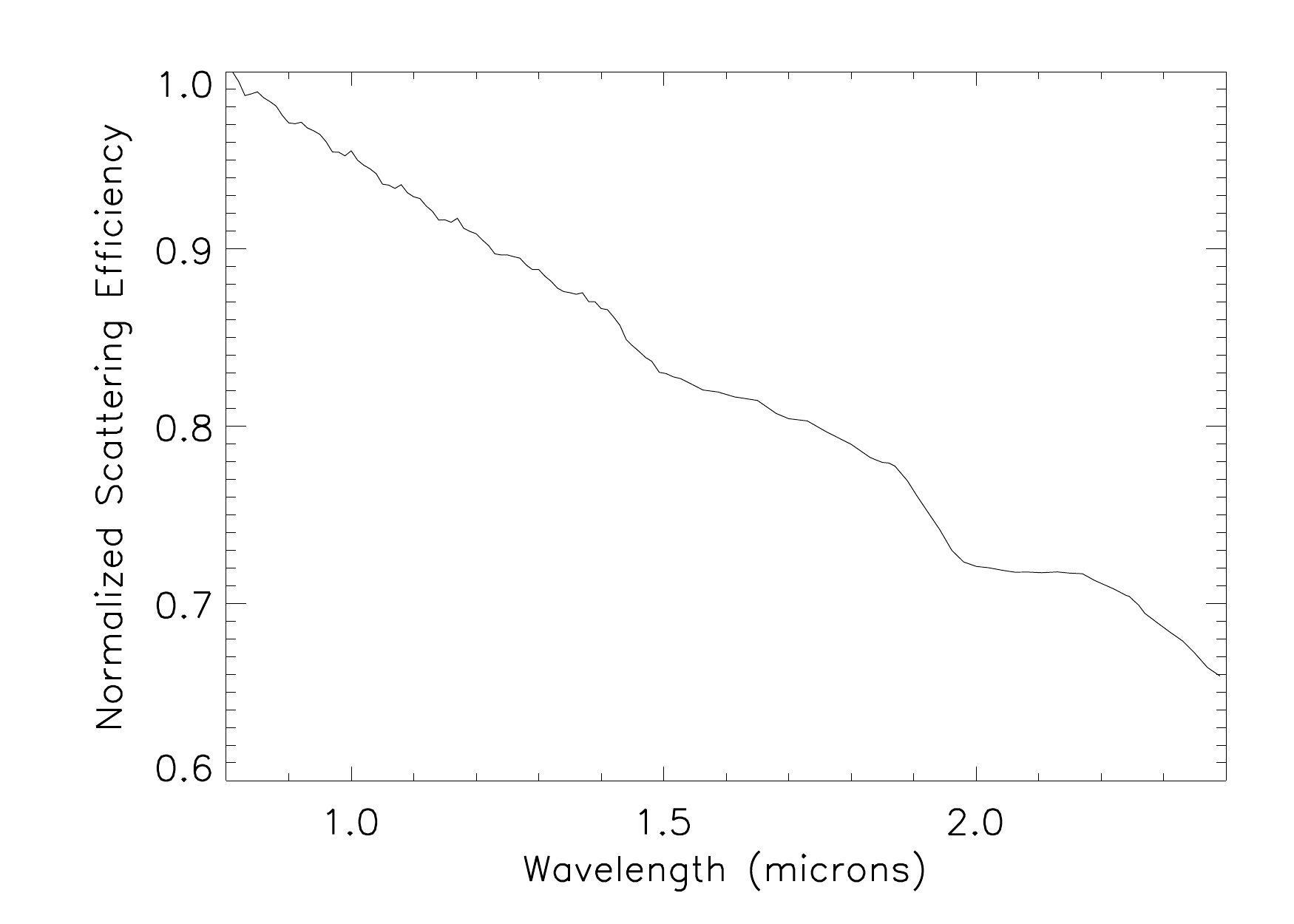}
\caption{We require a population of both small and large grains to reproduce the slope and water ice absorption features in the KH 15D reflectance spectrum. The most agreeable log-normal particle size distribution peaks at 1 $\mu$m $\left( \sigma = 2.0 \right)$.}
\label{miescatteringlognormal}
\end{figure}

After some exploration, we found that a particle size distribution that works well is a log-normal distribution with a peak at 1 $\mu$m and a $\sigma$ of 2.0. The bulk scattering efficiency of spherical ice grains with that size distribution is shown in Figure \ref{miescatteringlognormal}. It reproduces the observed blue slope and 1.5 and 2.0 $\mu$m absorption features rather well. We note that most of the mass in such a distribution would be in large grains $>$20 $\mu$m, although there are many more small grains. A more detailed modeling effort is postponed until we can obtain a larger spectral range for the reflectance spectrum.  

The KH 15D spectrum also shows shallow absorptions near 1.04, 1.25, 1.55, and 2.02 $\mu$m (see Table \ref{reflectancefeatures}, Figure \ref{reflectancespectrum}, bottom panel), which are the locations of prominent near-infrared features of water ice (e.g., \citet{Clark1980}). All four absorptions have been measured in spectra of the larger Jovian and Saturnian satellites and Saturn's rings \citep{Clark1980, Cruikshank2005, Filacchione2012}. The exact positions, shape, and depth of these bands is a function of grain size, crystallinity, and impurities. We conclude from this exercise that both the slope and primary absorption features of the observed reflectance spectrum can be understood as scattering from water ice grains with a distribution of grain sizes from about 1-100 $\mu$m. In order to confirm the detection of water ice in KH 15D, observations should be extended out to $\sim$5 $\mu$m. This will enable measurements of the 3 $\mu$m ice absorption that was used to set a lower limit on the abundance of ice in Comet 17P/Holmes \citep{Yang2009}.  

In addition to the water ice absorptions, the KH 15D spectrum has spectral features that are characteristic of hydrocarbons \citep{Cloutis1989}, specifically methane ice \citep{QS1997, Clark2009, Grundy2002}. Sublimation temperatures for these two constituents are around 160 and 45 K \citep{DR2009}, implying that the temperature in the scattering zone is very cold. Based on its precession rate, the KH 15D circumbinary ring is centered at a distance of 3.9 AU from the binary center of mass \citep{A16}. Its full width is expected to be approximately the same as its radial distance, although this dimension is much less certain than the radial distance. We may infer that the ring stretches roughly from 2-6 AU from the central stars, whose combined luminosity are currently $\sim 1 \, L_{\odot}$. It is unknown whether the occulting edge of the ring is its inner edge, its outer edge, or somewhere in between \citep{Winn2006}. It is therefore plausible that the scattering zone of the circumbinary ring would be outside the water frost line, but not the methane frost line, unless the grains are shielded from the direct radiation of the central stars.

\citet{DR2009} have explored models of ice line evolution in young disks that may be applicable to our observations. They find that by the age of the KH 15D system ($\sim$2 Myr) the methane ice line has moved in to about 5 AU, while the water ice line is at about 1 AU. Their model predicts that near 6 AU, coincident with the inferred position of the outer edge of the KH 15D circumbinary ring, the two most abundant ice species, by far, would be water and methane, which is just what we observe. It is uncertain how much relevance their model has, since it is for a fairly massive disk around a single star of 0.95 $M_{\odot}$, which is less than the combined mass of the KH 15D binary $\left(1.3 M_{\odot} \right)$. It also does not include the effect of settling by large dust grains, which may increase the shielding, providing a colder environment even closer to the central stars. 

To summarize, it is not surprising to find water ice as a principal component of the scattering grains in KH 15D, given the likely distance of the scattering zone from the stars, but it is somewhat surprising to find methane ice. While it could form and survive at 5 AU or even closer if shielded from the central starlight, it must be exposed to that starlight in order to actually scatter it. One would expect rapid sublimation of such ice exposed to the stellar radiation. As noted above, the required log-normal distribution of grain sizes suggests that most of the mass of the ice is in large grains, even pebbles or boulders. More sophisticated models of disk evolution may be required to account for our observation that water and methane ice are the dominant species in the KH 15D circumbinary ring. 

\begin{deluxetable}{cccc}
\tablewidth{0 pt}
\tabletypesize{\scriptsize}
\tablecaption{Reflectance Features of Dust in the Circumbinary Ring \label{reflectancefeatures}}
\tablehead{
\colhead{Water Ice} & \colhead{Methane Ice} & \colhead{Other Alkanes} & \colhead{Organic Solids} \\ 
\colhead{($\mu$m)} & \colhead{($\mu$m)} & \colhead{($\mu$m)} & \colhead{($\mu$m)} \\
}
\startdata
1.045 & 0.952 & 1.218 & 2.164 \\
1.256 & 1.003 & 2.281 & \\
1.486 & 1.160 & 2.343 & \\
1.547 & 1.670 & & \\
2.020 & 1.724 & & \\
 & 2.383 & & \\
\enddata
\end{deluxetable}

\subsection{Putative Giant Planet}

The GNIRS data were obtained at various orbital phases with the goal of isolating spectral signatures of the putative planet from features of the stars. The best opportunity to accomplish this comes during the faint phase, when the contribution to the continuum spectrum from scattered starlight is at a minimum. An average continuum value was determined in three different regions of the faint and intermediate phase spectra (corresponding to the $J$, $H$, and $K$ bands), and scale factors were determined for each region by dividing the continuum values from the faint spectra by those from the intermediate spectrum. The intermediate spectrum was then multiplied by the appropriate scale factors and subtracted from each of the two faint phase spectra in order to remove any residual features from the stars and the ring. 

\begin{figure}
\centering
\includegraphics[width=0.9\linewidth]
{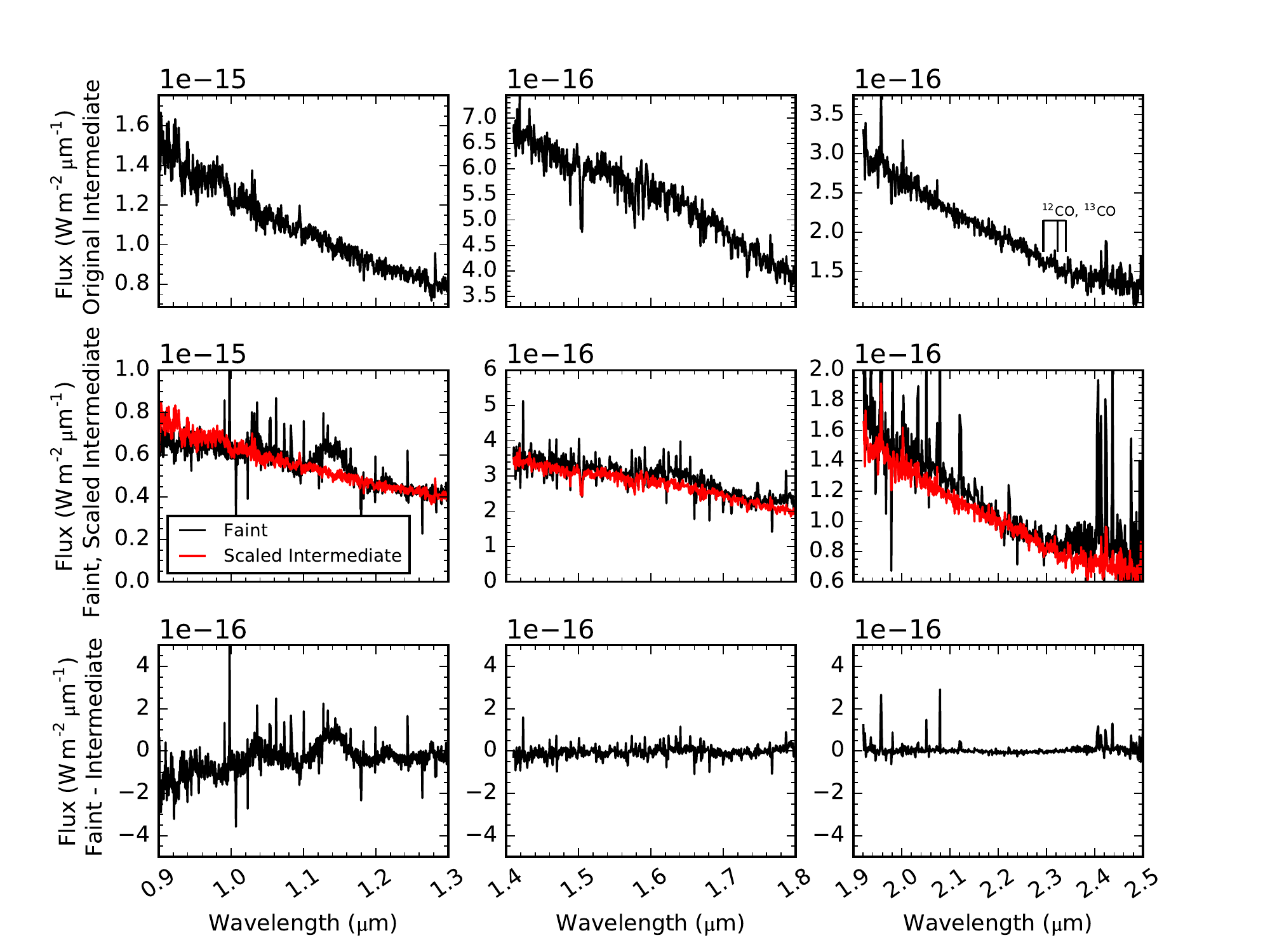}
\caption{The three rows illustrate the process of subtracting residual stellar and ring features from the November faint phase spectrum. The top row shows the original intermediate phase spectrum, which was scaled down to the flux level of the faint spectrum. The faint and scaled intermediate spectra are compared in the middle row, which shows that there is a small over-subtraction at shorter wavelengths. The scaled intermediate spectrum was then subtracted from the faint spectrum (bottom row).}
\label{novemberstarsubtraction}
\end{figure}

\begin{figure}
\centering
\includegraphics[width=1.0\linewidth]
{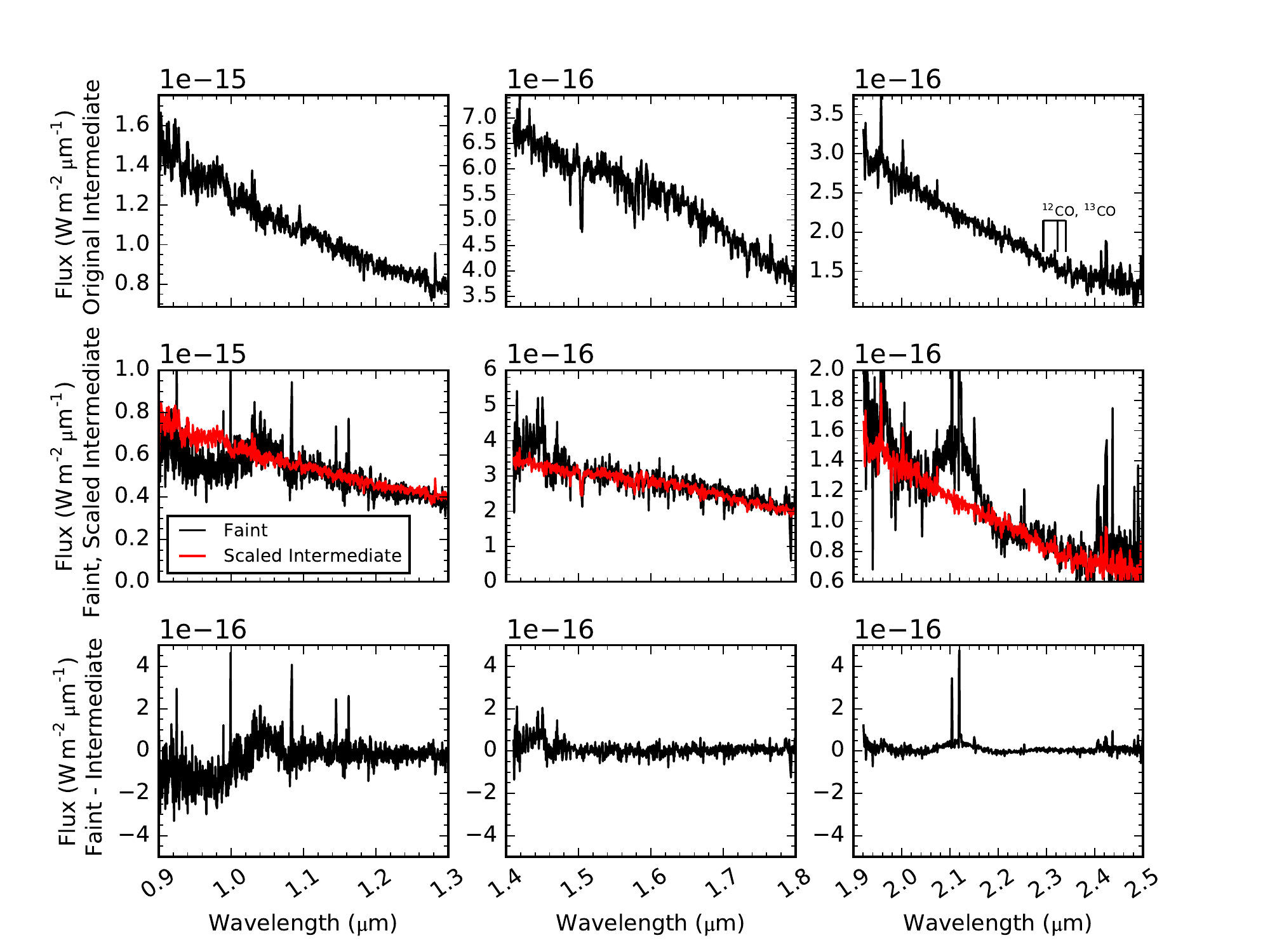}
\caption{The same procedure that was applied to the November faint phase spectrum (see Figure \ref{novemberstarsubtraction} was also applied to the December faint phase spectrum. As in Figure \ref{novemberstarsubtraction}, the three rows illustrate the star/ring subtraction process, here for the December data.}
\label{decemberstarsubtraction}
\end{figure}

\begin{figure}
\centering
\includegraphics[width=1.0\linewidth]
{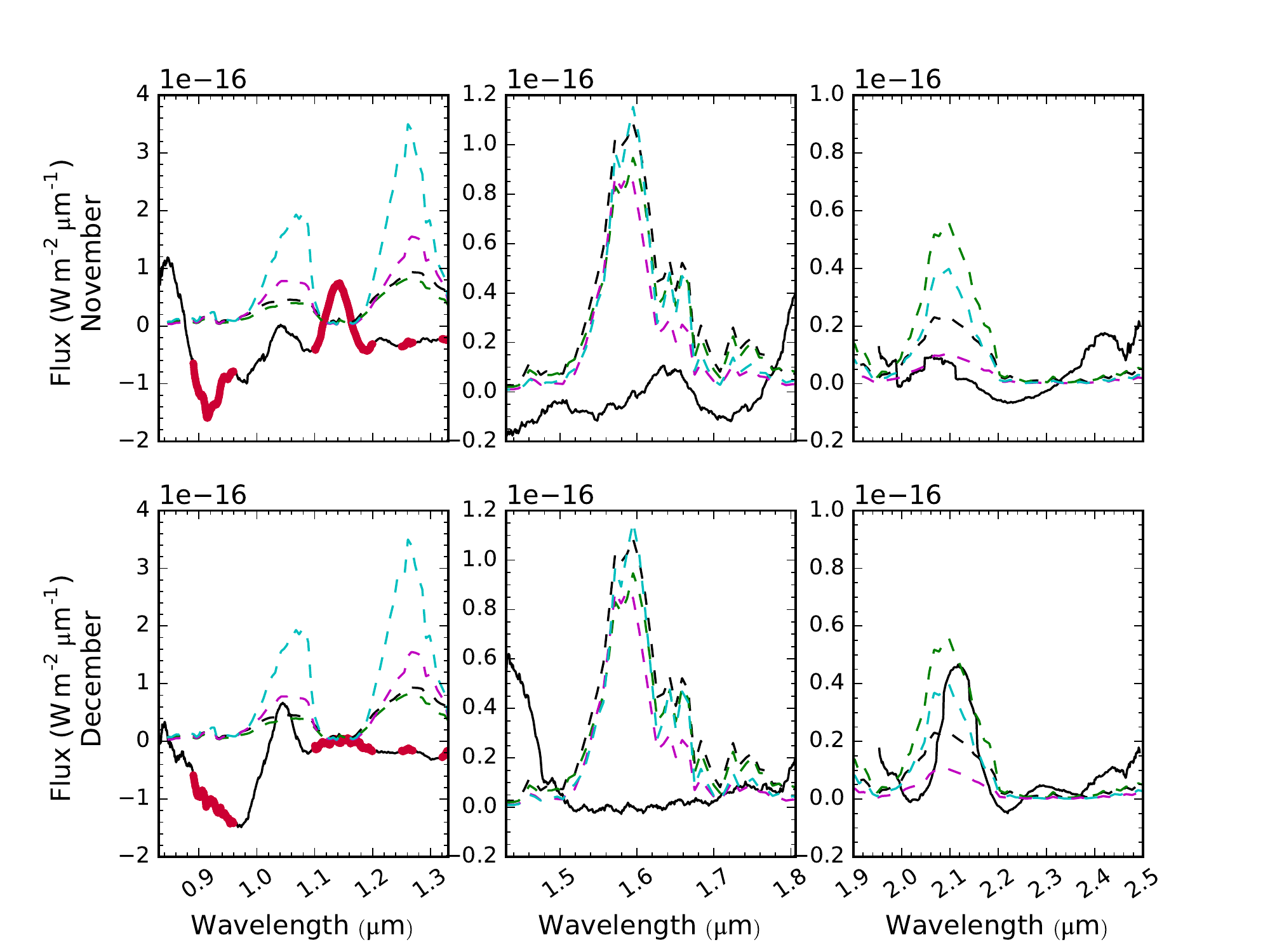}
\caption{We have compared the KH 15D spectra (solid, black) to model spectra of a 1 Myr, 10 $M_J$ planet from \cite{Spiegel2012} (dashed, colored). Each model spectrum assumes a different type of planetary atmosphere. Although we see a significant emission peak in the KH 15D data near the $\sim$2 $\mu$m NH$_3$ bump in the model spectra, the remaining features do not agree. The red regions from $\sim$0.9-1.35 $\mu$m in both spectra are locations of telluric absorptions, which were difficult to remove entirely from the low S/N data.}
\label{modelplanetspectra}
\end{figure}

Figures \ref{novemberstarsubtraction} and \ref{decemberstarsubtraction} illustrate the accuracy of this method by zooming in on three different regions of the faint phase spectra and comparing them to the scaled intermediate spectrum in the same wavelength ranges. A 50 point moving average was applied to the star-subtracted faint phase spectra in order to increase signal-to-noise after the prominent emission lines were removed. The resulting faint phase ``planet" spectra are plotted in Figure \ref{modelplanetspectra}. 

The scale factors that were applied to the intermediate spectrum have, for the most part, reduced its flux level to those of the faint spectra, although there is a small over-correction at shorter wavelengths that results in negative flux. This procedure could be improved by using a detailed model to predict what the scattered light spectrum will look like during the faint phase. It is also possible that the negative flux is due to variable telluric absorption, which is difficult to correct for in low S/N data. The telluric features are expected to be present at 0.89-0.96 $\mu$m, 1.11-1.17 $\mu$m, 1.25-1.27 $\mu$m, and 1.32-1.5 $\mu$m. These regions have been plotted in red in Figure \ref{modelplanetspectra}, and we caution the reader against over-interpreting the planet spectra at these wavelengths. 

Each faint spectrum was compared to model spectra of young giant planets from \cite{Spiegel2012} to see if any continuum features were similar to what is expected for a 1 Myr, 10 $M_J$ planet (see Figure \ref{modelplanetspectra}). The December spectrum in particular has a significant peak in emission near $\sim$2 $\mu$m and smaller bumps at the locations of other peaks in the model spectra. The features at 1.6 and 2.1 $\mu$m agree with model spectra of NH$_3$ and/or CH$_4$ \citep{Sharp2007}, while the features near 1.0 and 1.3 $\mu$m, if they are real, could be due either to H$_2$O or CH$_4$. As described in Section 3.4, detailed models of scattering and absorption signatures from the minerals we have detected in the circumbinary ring will require extending our observations out to $\sim$5 $\mu$m. Such models will be crucial to ensuring that the putative planet spectra can be accurately separated from the other components in the system. 

It is also possible that the increased reddening observed near minimum light is due to emission from dust in the ring, rather than a giant planet. However, \citet{Windemuth2014} determined that the amount of excess $H$ band flux was consistent with a $\sim$1000-2000 K source. Our analysis of the scattering properties of the circumbinary ring indicates that the dust temperature is much colder than even 1000 K, but longer wavelength observations are required to constrain this.

\section{Summary \& Conclusions}

We have analyzed spectra of KH 15D from three different orbital phases in order to untangle features from the five different components of the system that are detectable in its current geometry: 

\begin{enumerate}

\item \emph{Photosphere of Star B:} Absorption lines from $^{12}$CO, Si I, Na I, Fe I, Ca I, and Mg I were measured, allowing us to confirm the stellar spectral type. This absorption was traced through the intermediate and faint phases.

\item \emph{Magnetosphere of Star B:} He I $\lambda$10830 emission was observed at all three phases, and a red-shifted absorption component was detected during the bright phase. The line fluxes in the four spectra were inconsistent with each other, and we suspect that heating events may be the cause of this variability. 

\item \emph{Inner Jet:} We measured $H_2$ emission lines in the intermediate and faint phase spectra and derived a thermal excitation temperature for each data set. The three results are consistent and also agree with measurements by \citet{Deming2004}, which confirms that the features originate at the base of a bipolar outflow that has remained unocculted as the system's geometry has evolved. 

\item \emph{Circumbinary Ring:} Reflectance features from material in the ring were studied by dividing the bright spectrum by the spectrum from the intermediate phase, and we have identified absorptions from water ice and methane. A mixture of grains that peaks at 1 $\mu$m and contains particles up to 100 $\mu$m in size is required to reproduce the observed blue scattering slope and ice absorptions.

\item \emph{Putative Planet:} After removing the features that were identified in the other four components of KH 15D, we detected a prominent leftover peak in emission that is consistent with model spectra of a 1 Myr, 10 $M_J$ planet. However, the remaining features in this putative planet spectrum are unaccounted for. 

\end{enumerate}

The unique geometry of KH 15D has allowed us to try a new method of directly observing a planet. Although our results do not confirm the presence of a shepherding body at the outer edge of the system's circumbinary ring, it is possible that the method may be more successful for objects that are closer to Earth. A key source of uncertainty was our removal of reflectance features and scattered light. Since the resolution of our data are low, it is difficult to determine the detailed composition of the material that produced this reflectance spectrum and predict how it will behave when the system is faint. 

It is not currently feasible to obtain spatially resolved spectra of KH 15D, but there are many nearby disks around young stars that produce strong near-infrared signals in scattered light (e.g. \citet{Debes2013}, \citet{Avenhaus2014B}, \citet{Avenhaus2014A}). \citet{Honda2016} were able to use scattered light imaging to detect a 3.1 $\mu$m water ice feature in the disk of HD 100546, therefore confirming that scattered light photons contain detectable signatures of the disk material. Reflectance spectra similar to the one we produced for KH 15D can be acquired for other systems by dividing the scattered light by templates corresponding to the host stars' spectral type \citep{Debes2013}. This would produce spatially resolved reflectance spectra, allowing us to carry out a more detailed search for prominent near-infrared features from minerals in the disk as a function of distance from the central star.

\acknowledgments
This work is based on observations at the Gemini Observatory, under Program ID GN-2013A-Q-27. The Gemini Observatory is operated by the Association of Universities for Research in Astronomy, Inc., under a cooperative agreement with the NSF on behalf of the Gemini partnership: the National Science Foundation (United States), the Particle Physics and Astronomy Research Council (United Kingdom), the National Research Council (Canada), CONICYT (Chile), the Australian Research Council (Australia), CNPq (Brazil) and CONICET (Argentina). This research made use of Astropy, a community-developed core Python package for Astronomy \citep{Astropy2013}.

\bibliographystyle{apj}
\bibliography{gemini_bibliography}

\begin{thebibliography}{38}
\expandafter\ifx\csname natexlab\endcsname\relax\def\natexlab#1{#1}\fi

\bibitem[{{Agol} {et~al.}(2004){Agol}, {Barth}, {Wolf}, \&
  {Charbonneau}}]{Agol2004}
{Agol}, E., {Barth}, A.~J., {Wolf}, S., \& {Charbonneau}, D. 2004, \apj, 600,
  781

\bibitem[{{Arulanantham} {et~al.}(2016){Arulanantham}, {Herbst}, {Cody},
  {Stauffer}, {Rebull}, {Agol}, {Windemuth}, {Marengo}, {Winn}, {Hamilton},
  {Mundt}, {Johns-Krull}, \& {Gutermuth}}]{A16}
{Arulanantham}, N.~A., {et~al.} 2016, \aj, 151, 90

\bibitem[{{Astropy Collaboration} {et~al.}(2013){Astropy Collaboration},
  {Robitaille}, {Tollerud}, {Greenfield}, {Droettboom}, {Bray}, {Aldcroft},
  {Davis}, {Ginsburg}, {Price-Whelan}, {Kerzendorf}, {Conley}, {Crighton},
  {Barbary}, {Muna}, {Ferguson}, {Grollier}, {Parikh}, {Nair}, {Unther},
  {Deil}, {Woillez}, {Conseil}, {Kramer}, {Turner}, {Singer}, {Fox}, {Weaver},
  {Zabalza}, {Edwards}, {Azalee Bostroem}, {Burke}, {Casey}, {Crawford},
  {Dencheva}, {Ely}, {Jenness}, {Labrie}, {Lim}, {Pierfederici}, {Pontzen},
  {Ptak}, {Refsdal}, {Servillat}, \& {Streicher}}]{Astropy2013}
{Astropy Collaboration} {et~al.} 2013, \aap, 558, A33

\bibitem[{{Avenhaus} {et~al.}(2014{\natexlab{a}}){Avenhaus}, {Quanz}, {Meyer},
  {Brittain}, {Carr}, \& {Najita}}]{Avenhaus2014B}
{Avenhaus}, H., {Quanz}, S.~P., {Meyer}, M.~R., {Brittain}, S.~D., {Carr},
  J.~S., \& {Najita}, J.~R. 2014{\natexlab{a}}, \apj, 790, 56

\bibitem[{{Avenhaus} {et~al.}(2014{\natexlab{b}}){Avenhaus}, {Quanz}, {Schmid},
  {Meyer}, {Garufi}, {Wolf}, \& {Dominik}}]{Avenhaus2014A}
{Avenhaus}, H., {Quanz}, S.~P., {Schmid}, H.~M., {Meyer}, M.~R., {Garufi}, A.,
  {Wolf}, S., \& {Dominik}, C. 2014{\natexlab{b}}, \apj, 781, 87

\bibitem[{{Capelo} {et~al.}(2012){Capelo}, {Herbst}, {Leggett}, {Hamilton}, \&
  {Johnson}}]{Capelo2012}
{Capelo}, H.~L., {Herbst}, W., {Leggett}, S.~K., {Hamilton}, C.~M., \&
  {Johnson}, J.~A. 2012, \apjl, 757, L18

\bibitem[{{Chiang} \& {Murray-Clay}(2004)}]{CM2004}
{Chiang}, E.~I., \& {Murray-Clay}, R.~A. 2004, \apj, 607, 913

\bibitem[{{Clark} {et~al.}(2009){Clark}, {Curchin}, {Hoefen}, \&
  {Swayze}}]{Clark2009}
{Clark}, R.~N., {Curchin}, J.~M., {Hoefen}, T.~M., \& {Swayze}, G.~A. 2009,
  Journal of Geophysical Research (Planets), 114, E03001

\bibitem[{{Clark} \& {McCord}(1980)}]{Clark1980}
{Clark}, R.~N., \& {McCord}, T.~B. 1980, \icarus, 43, 161

\bibitem[{{Cloutis}(1989)}]{Cloutis1989}
{Cloutis}, E.~A. 1989, Science, 245, 165

\bibitem[{{Cruikshank} {et~al.}(2005){Cruikshank}, {Owen}, {Dalle Ore},
  {Geballe}, {Roush}, {de Bergh}, {Sandford}, {Poulet}, {Benedix}, \&
  {Emery}}]{Cruikshank2005}
{Cruikshank}, D.~P., {et~al.} 2005, \icarus, 175, 268

\bibitem[{{Dabrowski}(1984)}]{Dabrowski1984}
{Dabrowski}, I. 1984, Canadian Journal of Physics, 62, 1639

\bibitem[{{Debes} {et~al.}(2013){Debes}, {Jang-Condell}, {Weinberger},
  {Roberge}, \& {Schneider}}]{Debes2013}
{Debes}, J.~H., {Jang-Condell}, H., {Weinberger}, A.~J., {Roberge}, A., \&
  {Schneider}, G. 2013, \apj, 771, 45

\bibitem[{{Deming} {et~al.}(2004){Deming}, {Charbonneau}, \&
  {Harrington}}]{Deming2004}
{Deming}, D., {Charbonneau}, D., \& {Harrington}, J. 2004, \apjl, 601, L87

\bibitem[{{Dodson-Robinson} {et~al.}(2009){Dodson-Robinson}, {Willacy},
  {Bodenheimer}, {Turner}, \& {Beichman}}]{DR2009}
{Dodson-Robinson}, S.~E., {Willacy}, K., {Bodenheimer}, P., {Turner}, N.~J., \&
  {Beichman}, C.~A. 2009, \icarus, 200, 672

\bibitem[{{Filacchione} {et~al.}(2012){Filacchione}, {Capaccioni},
  {Ciarniello}, {Clark}, {Cuzzi}, {Nicholson}, {Cruikshank}, {Hedman},
  {Buratti}, {Lunine}, {Soderblom}, {Tosi}, {Cerroni}, {Brown}, {McCord},
  {Jaumann}, {Stephan}, {Baines}, \& {Flamini}}]{Filacchione2012}
{Filacchione}, G., {et~al.} 2012, \icarus, 220, 1064

\bibitem[{{Fischer} {et~al.}(2008){Fischer}, {Kwan}, {Edwards}, \&
  {Hillenbrand}}]{Fischer2008}
{Fischer}, W., {Kwan}, J., {Edwards}, S., \& {Hillenbrand}, L. 2008, \apj, 687,
  1117

\bibitem[{{F{\"o}rster Schreiber}(2000)}]{FS2000}
{F{\"o}rster Schreiber}, N.~M. 2000, \aj, 120, 2089

\bibitem[{{Grundy} {et~al.}(2002){Grundy}, {Schmitt}, \&
  {Quirico}}]{Grundy2002}
{Grundy}, W.~M., {Schmitt}, B., \& {Quirico}, E. 2002, \icarus, 155, 486

\bibitem[{{Hamilton} {et~al.}(2001){Hamilton}, {Herbst}, {Shih}, \&
  {Ferro}}]{Hamilton2001}
{Hamilton}, C.~M., {Herbst}, W., {Shih}, C., \& {Ferro}, A.~J. 2001, \apjl,
  554, L201

\bibitem[{{Hamilton} {et~al.}(2005){Hamilton}, {Herbst}, {Vrba}, {Ibrahimov},
  {Mundt}, {Bailer-Jones}, {Filippenko}, {Li}, {B{\'e}jar}, {{\'A}brah{\'a}m},
  {Kun}, {Mo{\'o}r}, {Benk{\H o}}, {Csizmadia}, {DePoy}, {Pogge}, \&
  {Marshall}}]{Hamilton2005}
{Hamilton}, C.~M., {et~al.} 2005, \aj, 130, 1896

\bibitem[{{Honda} {et~al.}(2016){Honda}, {Kudo}, {Takatsuki}, {Inoue},
  {Nakamoto}, {Fukagawa}, {Tamura}, {Terada}, \& {Takato}}]{Honda2016}
{Honda}, M., {et~al.} 2016, \apj, 821, 2

\bibitem[{{Ishii} {et~al.}(2004){Ishii}, {Tamura}, \& {Itoh}}]{IT2004}
{Ishii}, M., {Tamura}, M., \& {Itoh}, Y. 2004, \apj, 612, 956

\bibitem[{{Kearns} \& {Herbst}(1998)}]{Kearns1998}
{Kearns}, K.~E., \& {Herbst}, W. 1998, \aj, 116, 261

\bibitem[{{Quirico} \& {Schmitt}(1997)}]{QS1997}
{Quirico}, E., \& {Schmitt}, B. 1997, \icarus, 127, 354

\bibitem[{{Sharp} \& {Burrows}(2007)}]{Sharp2007}
{Sharp}, C.~M., \& {Burrows}, A. 2007, \apjs, 168, 140

\bibitem[{{Silvia} \& {Agol}(2008)}]{Silvia2008}
{Silvia}, D.~W., \& {Agol}, E. 2008, \apj, 681, 1377

\bibitem[{{Skrutskie} {et~al.}(2006){Skrutskie}, {Cutri}, {Stiening},
  {Weinberg}, {Schneider}, {Carpenter}, {Beichman}, {Capps}, {Chester},
  {Elias}, {Huchra}, {Liebert}, {Lonsdale}, {Monet}, {Price}, {Seitzer},
  {Jarrett}, {Kirkpatrick}, {Gizis}, {Howard}, {Evans}, {Fowler}, {Fullmer},
  {Hurt}, {Light}, {Kopan}, {Marsh}, {McCallon}, {Tam}, {Van Dyk}, \&
  {Wheelock}}]{2MASS}
{Skrutskie}, M.~F., {et~al.} 2006, \aj, 131, 1163

\bibitem[{{Spiegel} \& {Burrows}(2012)}]{Spiegel2012}
{Spiegel}, D.~S., \& {Burrows}, A. 2012, \apj, 745, 174

\bibitem[{{Tokunaga} {et~al.}(2004){Tokunaga}, {Dahm}, {G{\"a}ssler}, {Hayano},
  {Hayashi}, {Iye}, {Kanzawa}, {Kobayashi}, {Kamata}, {Minowa}, {Nedachi},
  {Oya}, {Pyo}, {Saint-Jacques}, {Terada}, {Takami}, \&
  {Takato}}]{Tokunaga2004}
{Tokunaga}, A.~T., {et~al.} 2004, \apjl, 601, L91

\bibitem[{{Turner} {et~al.}(1977){Turner}, {Kirby-Docken}, \&
  {Dalgarno}}]{Turner1977}
{Turner}, J., {Kirby-Docken}, K., \& {Dalgarno}, A. 1977, \apjs, 35, 281

\bibitem[{{van Belle} {et~al.}(1999){van Belle}, {Lane}, {Thompson}, {Boden},
  {Colavita}, {Dumont}, {Mobley}, {Palmer}, {Shao}, {Vasisht}, {Wallace},
  {Creech-Eakman}, {Koresko}, {Kulkarni}, {Pan}, \& {Gubler}}]{VB1999}
{van Belle}, G.~T., {et~al.} 1999, \aj, 117, 521

\bibitem[{{Vilas} {et~al.}(1984){Vilas}, {Leake}, \& {Mendell}}]{Vilas1984}
{Vilas}, F., {Leake}, M.~A., \& {Mendell}, W.~W. 1984, \icarus, 59, 60

\bibitem[{{Warren}(1984)}]{Warren1984}
{Warren}, S.~G. 1984, \ao, 23, 1206

\bibitem[{{Windemuth} \& {Herbst}(2014)}]{Windemuth2014}
{Windemuth}, D., \& {Herbst}, W. 2014, \aj, 147, 9

\bibitem[{{Winn} {et~al.}(2006){Winn}, {Hamilton}, {Herbst}, {Hoffman},
  {Holman}, {Johnson}, \& {Kuchner}}]{Winn2006}
{Winn}, J.~N., {Hamilton}, C.~M., {Herbst}, W.~J., {Hoffman}, J.~L., {Holman},
  M.~J., {Johnson}, J.~A., \& {Kuchner}, M.~J. 2006, \apj, 644, 510

\bibitem[{{Yang} {et~al.}(2009){Yang}, {Jewitt}, \& {Bus}}]{Yang2009}
{Yang}, B., {Jewitt}, D., \& {Bus}, S.~J. 2009, \aj, 137, 4538

\bibitem[{{Zeng} {et~al.}(2014){Zeng}, {Qiu}, {Cao}, \& {Judge}}]{Zeng2014}
{Zeng}, Z., {Qiu}, J., {Cao}, W., \& {Judge}, P.~G. 2014, \apj, 793, 87

\end{thebibliography}

\end{document}